\documentclass[fleqn,usenatbib]{mnras}

\usepackage{newtxtext,newtxmath}
 
\usepackage[T1]{fontenc}

\DeclareRobustCommand{\VAN}[3]{#2}
\let\VANthebibliography\thebibliography
\def\thebibliography{\DeclareRobustCommand{\VAN}[3]{##3}\VANthebibliography}

\usepackage{graphicx}	
\usepackage{amsmath}	
\usepackage{amssymb}	
\usepackage{xcolor}

\defcitealias{Rosotti2019}{R19}
\defcitealias{Rosotti2019L}{R19L}
\defcitealias{Trapman2019}{T19}
\defcitealias{Trapman2020}{T20}

\title[Efficiency of radial drift]
{On the secular evolution of the ratio between gas and dust radii in protoplanetary discs}

\author[C. Toci et al.]{
Claudia Toci$^{1}$\thanks{E-mail: claudia.toci@unimi.it},
Giovanni Rosotti$^{2,3}$,
Giuseppe Lodato$^{1}$,
Leonardo Testi$^{4}$,
Leon Trapman$^{5,3}$
\\
$^{1}$Dipartimento di Fisica, Università degli Studi di Milano, Via Giovanni Celoria, 16, 20133 Milano, MI, Italy \\
$^{2}$School of Physics and Astronomy, University of Leicester, Leicester LE1 7RH, UK\\
$^{3}$Leiden University, Niels Bohrweg 2, NL-2333 CA Leiden, The Netherlands\\
$^{4}$European Southern Observatory (ESO), Karl-Schwarzschild-Str. 2, 85748, Garching, Germany \\
$^{5}$Department of Astronomy, University of Wisconsin-Madison, 475 N Charter St, Madison, WI, USA\\
}

\date{Accepted XXX. Received YYY; in original form ZZZ}

\pubyear{2021}

\begin{document}
\label{firstpage}
\pagerange{\pageref{firstpage}--\pageref{lastpage}}
\maketitle

\begin{abstract}
A key problem in protoplanetary disc evolution is
understanding the efficiency of dust radial drift. This process makes the observed dust disc sizes shrink on relatively short timescales, implying that discs started much larger than what we see now. In this paper we use an independent constraint, the gas radius (as probed by CO rotational emission), to test disc evolution models. In particular, we consider the ratio between the dust and gas radius, $R_{\rm CO}/R_{\rm dust}$. We model the time evolution of protoplanetary discs under the influence of viscous evolution, grain growth, and radial drift. Then, using the radiative transfer code RADMC with approximate chemistry, we compute the dust and gas radii of the models and investigate how $R_{\rm CO}/R_{\rm dust}$ evolves. Our main finding is that, for a broad range of values of disc mass, initial radius, and viscosity, $R_{\rm CO}/R_{\rm dust}$ becomes large (>5)
after only a short time (<1 Myr) due to radial drift. This is at odds with measurements in young star forming regions such as Lupus, which find much smaller values, implying that dust radial drift is too efficient in these models. 
Substructures, commonly invoked to stop radial drift in large, bright discs, must then be present, although currently unresolved, in most discs.
\end{abstract}

\begin{keywords}
accretion, accretion discs -- planets and satellites: formation -- protoplanetary discs
\end{keywords}



\section{Introduction}
How protoplanetary discs evolve has a huge impact on the final architecture \citep{Ford2014} and composition of the resulting planetary system \citep{mordasini2015}: while young protoplanets are forming and accreting, embedded in the parental environment, the surrounding material continuously changes in physical conditions and composition \citep{Testi2014,Manara2019}.

One of the biggest open problems in disc evolution is understanding the role of dust radial drift \citep{Weidenschilling1977}. As a result of the azimuthal headwind from the gas slowing down dust particles, large grains radially drift inward on a timescale much shorter than the lifetime of the disc. As a consequence, the dust disc size shrinks in time and eventually, due to the very short timescales involved, the dust disc disappears \citep{Takeuchi2005,Appelgren2020}. Small grains that are left behind in the disc can in principle be observed for longer timescales (several Myr). However, such discs are expected to be large and faint \citep{Rosotti2019}, thus they cannot represent the bulk of the observed disc population.
It is however an observational fact that discs are still observed at a typical age of a few Myr. A possible way to explain why this is the case is offered by dust traps, collecting the dust at pressure maxima and in this way slowing down or stopping radial drift \citep{Whipple1972,Pinilla2012}. 

Thanks to the new telescopes, pressure traps are routinely observed. Indeed, with the Atacama Large Millimetre Array (ALMA) and the Spectro-Polarimetric High-contrast Exoplanet REsearch (SPHERE) instrument at the Very Large Telescope (VLT), we now have high resolution (<0.05\arcsec) images of large ( > 50 au) and bright discs. Many of these present substructures such as rings, gaps, and spirals \citep{ALMA2015,Andrews2018,Andrews2020}, directly or speculatively connected to the presence of young planets (e.g., \citealt{Mueller2018, Pinte2019,Lodato19}). 
Rings are of particular interest because they likely constitute dust traps \citep{Dipierro2015, Dullemond2018,Rosotti2020}, explaining why the dust discs have not drained onto the star due to radial drift. 
However, bright and large discs are not the majority among the disc population, and could give a very biased view: as an example, the faintest disc imaged by the DSHARP survey \citep{Andrews2018} in Lupus has a 1.3 mm flux of 77 mJy, corresponding to the top $\sim 5$\% of the disc luminosity function in Lupus \citep{Ansdell2016,Ansdell2018}.

As a complementary point of view, it is also possible to perform entire surveys of a star forming region, allowing us also to investigate how disc properties vary with age. For example, dust disc masses with ALMA and mass accretion rates with spectroscopic observations have been measured in Chamaeleon I \citep{Manara2016,Pascucci2016}, Lupus \citep{Ansdell2016,Alcala2017,Tazzari2020, Hendler2020,Sanchis2020,Sanchis2021} and Upper Scorpius \citep{Barenfeld2017,Manara2020}, and dust disc masses are available in many other regions \citep[e.g.,][]{Cieza2019,Cazzoletti2019}. While observational time constraints in this case do not allow high spatial resolution, these studies have a much larger statistics that can be used to test theories of disc evolution \citep{Rosotti2017,Lodato2017,Mulders2017,Ercolano2017}, finding that in at least two regions -- Lupus and Chamaeleon -- there is a correlation between the two quantities, as predicted for viscous evolution models \citep{Lodato2017}. Photoevaporation is possibly at play in the older region Upper Scorpius \citep{Somigliana2020}. 

Many authors measured also the extent of the mm thermal continuum emission, i.e. the dust disk size (see e.g., \citealt{Cox2017, Cieza2019} for Ophiucus, \citealt{Ansdell2016,Tazzari2017,Tazzari2020,Hendler2020,Sanchis2021} for Lupus,   \citealt{Long2019, Kurtovic2021} for Taurus). 
Unfortunately, for most of these discs we do not know if they possess substructures. For simplicity, \citet{Rosotti2019} (hereafter \citetalias{Rosotti2019}) modelled them as smooth. They found that it is still possible to explain the observed fluxes and dust radii of discs in models that include radial drift, despite its fast timescale. Their explanation is that radial drift is "a victim of its own success" and that by removing the largest grains it leaves behind the slowest drifting grains. As shown by \citet{Rosotti2019L} (hereafter \citetalias{Rosotti2019L}), not only radial drift does not cause the dust disc to disappear, but even explains the observed correlation between dust disc size and flux (see e.g \citealt{Thripati2018}). Because only the grains larger than the wavelength have significant opacity to be observed, the physics of grain growth and drift is fundamental in setting the observed disc size.
Moreover, \citet{Sellek2020} show that smooth models including radial drift can also explain the distribution of observations in the accretion rate - disc mass plane, even in the old Upper Sco region.

The question on the role of radial drift, and as a consequence the role of substructures, thus remains open. The success of smooth models in interpreting the bulk of the disc population could lead us to believe that, if interested in radial drift, substructures are not important after all for the majority of discs, despite being commonly observed in bright discs. We could even speculate that the large, bright discs are such \textit{because} they posses substructure which prevented radial drift, while most discs are smooth and therefore became fainter and smaller (see e.g., discussions in \citealt{Long2019} and \citealt{Banzatti2020}). One worrying sign that this may not be the case is offered by the study of dust spectral indices. Because radial drift removes the largest grains, as time progresses smooth models contain relatively small grains, which observationally have large spectral indices \citep{Birnstiel2010b}. This is not compatible with observations, and while until recently spectral indices had been measured only for large, bright discs \citep{Ricci2010a,Ricci2010b}, ALMA is showing us that this also found for the bulk of the disc population \citep{Tazzari2020} (however, \citet{Tazzari2020} focused on the brightest discs in Lupus, about $50\%$ of the population.). Models including dust trap do not suffer from having large spectral indices \citep{Pinilla2012}, but fail in explaining the observed properties (e.g., \citealt{Pinilla2020}).

Ultimately, the problem is that these smooth models have been constructed only to reproduce the dust size. Here the models have quite some freedom because they can assume that the true extent of the disc is much larger than the observed dust extent, due to the shrinking effect of radial drift. There is however another observational quantity that should be able to settle this question: the gas disc size (hereafter called $R_{\rm CO}$). This is commonly measured using the rotational line emission of the CO molecule since it is by far the brightest emission line in proto-planetary discs. Even so, observing CO is more time consuming than the continuum, and this becomes even more constraining for its less abundant isotopologues ($^{13}$CO and C$^{18}$O) which are better mass tracers due to the lower optical depth. Only in the last few years some regions have been surveyed for CO emission \citep{Barenfeld2017, Ansdell2018}, but many discs seen in the dust continuum still lack a detected gaseous counterpart. Indeed, recent studies found that undetected discs in CO and $^{13}$CO emission may be intrinsically compact, and can be more than 50$\%$ of the population of the Lupus region \citep{Miotello2021}. Chemical processes such as photo-dissociation by UV radiation in the upper layers and freeze-out onto grains in the disc midplane are at play, potentially complicating the interpretation \citep{Dutrey1997}. It is nevertheless possible to include these effects in the modelling \citep{Williams2014,Facchini2017,Trapman2019} - as shown by \citet{Trapman2019} (hereafter \citetalias{Trapman2019}), $R_{\rm CO}$ has a rather straightforward interpretation as the radius where CO is photodissociated.

In the same way as the dust size evolves, also the gas size is not constant with time. However, the physical process governing its evolution is different and it is linked to the mechanism driving accretion onto the star. We do not wish here to enter the current debate regarding whether accretion is driven by viscosity or MHD winds; in this paper we will simply assume that disc evolution is driven by viscosity. In this picture, the disc radius generally gets larger with time (the so called viscous spreading) due to the redistribution of angular momentum in the disc. This view is partially supported by \citet{Najita2018}, who suggested that the observed sizes of most Class II gas discs are larger compared to those of Class I, in agreement with viscous spreading, although with the significant caveat that different molecular tracers are used for the different classes. More recently, \citet{Trapman2020} (hereafter \citetalias{Trapman2020}) modelled the density structures of discs at different ages with the thermochemical code DALI \citep{Bruderer2012,Bruderer2014}.
They compared their $R_{\rm CO}$ predictions with an observational sample, which uniformly uses the CO tracer, from the young Lupus (1-3 Myr, \citealt{Ansdell2018}) and the older Upper Scorpius (5-10~Myr, \citealt{Barenfeld2017}) regions. No "smoking gun" signature of viscous spreading is observed, since the discs in Upper Sco are smaller than in Lupus, though this is likely to be an effect of external photo-evaporation.   

A way to test how much dust is drifting in discs is to analyse the secular evolution of the \textit{ratio} between dust and gas radii, $R_{\rm CO}/R_{\rm dust}$. In their latest work, \citealt{Sanchis2021} studied this quantity for Lupus, finding that the vast majority of the discs from his sample, about $\sim 50\%$ of the discs in Lupus, has a very similar size ratio (between 2 and 4), with a few sources ($15\%$ of the population), with size ratio > 4, suggesting that a large fraction of discs in Lupus evolve similarly and may be in a similar evolutionary stage. Since we observe the dust radius of discs at a single moment of their lifetime, we can explain them as a) the effect of radial drift on an initially very large disc or b) the presence of substructures that halted dust drift without modifying the disc size. By investigating the ratio of dust and gas radii, we break this degeneracy: we expect to find large gas radii in the former case and gas radii comparable to the dust in the latter case. 

To investigate this idea quantitatively, in this work we perform a follow up of \citetalias{Rosotti2019} and \citetalias{Rosotti2019L}, testing how much dust radially drift in synthetic populations of protoplanetary discs with respect to the gas.
We compute the expected $^{12}$CO radii associated to the models used by \citetalias{Rosotti2019} and \citetalias{Rosotti2019L} to compute the dust size, assuming viscous evolution, to study their time evolution and the secular evolution of the ratio $R_{\rm CO}/R_{\rm dust}$. We will assess whether smooth models, which do not take into account the possibility of trapping dust and forming substructures, are able to reproduce the observed range of $R_{\rm CO}/R_{\rm dust}$, and if so under which conditions.

The paper is organised as follow: In Sec.~\ref{Model} we introduce our model, discussing the methods and the assumption we chose. In Sec.~\ref{Results} we first describe in details our representative case, then we show the effects of the parameters on our results. We discuss our findings and their implications in Sec.~\ref{Discussion} and we give our conclusions in Sec.~\ref{concl}.

\section{Methods}\label{Model}

We first briefly summarise the most important aspects of the numerical code, presented in \citet{Booth2017} and already used in \citetalias{Rosotti2019} and \citetalias{Rosotti2019L},  then we report how we set up our grid of models and finally we describe how we calculate the surface density profiles and the dust and gas radii.

We set up a suite of 1D models of discs composed by gas and dust. The algorithm evolves the gas and two dust components at each radius: a small (0.1 $\mu$m size) population, larger in number, and a large (1 mm) one, dominant in mass.
The gas evolves following the viscous evolution equations, while dust growth is implemented according to the prescription of \citet{Birnstiel2012}, an accurate (at least for the smooth discs we model in this work) but computationally cheap approach compared with more expensive codes (see e.g \citealt{Birnstiel2010}). Considering dust growth and the consequent radial drift in the model results in the possibility to have spatially different values of the gas to dust ratio each timestep, which was instead kept fixed in previous works (see \citetalias{Trapman2019,Trapman2020}), allowing us to test the effect of radial drift on discs sizes.

We initialise our set of models with initial conditions that span different values of the parameter space of protoplanetary discs, focusing on Solar mass stars. We choose this approach because this paper is focused in identifying the physical mechanisms governing the evolution of the relative dust and gas radii; in the future, we plan to make more tailored comparisons to the available observations using a population synthesis approach. For each set of initial conditions, we evolve our populations from 0 to 3~Myr outputting the results every $10^5$ yr, creating synthetic emissions for the gas and the dust. 
The synthetic surface brightness profiles of mm continuum emissions and the dust radii values are evaluated as in \citetalias{Rosotti2019}. To produce synthetic observations of the CO rotational lines we performed radiative transfer simulations with the RADMC-3D code \citep{Dullemond2012}, measuring the radius enclosing 68$\%$ of the CO flux, that for this work (but see also \citetalias{Trapman2020}) represents the observed gas disc radii $R_{\rm CO}$. Similarly, the dust radius is evaluated as the radius enclosing 68$\%$ of the millimetric flux (see also \citetalias{Rosotti2019L}).

\subsection{Disc evolution}

The disc kinematic viscosity can be parameterized \citep{SS74} as $\nu=\alpha c_sH$, where $c_s$ is the sound speed and $H$ is the scale height of the disc. 
We assume a Solar mass star. The temperature of the disc is described by a time independent, radial power-law, as $T(R)=40 (R/\mbox{10au})^{-0.5}$ K, where the normalization is calibrated at 10 au using the \citet{Chiang1997} two-layers model. This corresponds to an aspect ratio of $H/R=0.033 ({ R/ 1\rm{au}})^{1/4}$. With this choice, and assuming a constant $\alpha$, the viscosity varies radially as $\nu \propto R/R_{\rm c}$, where $R_{\rm c}$ is the characteristic scale of the disc (containing (1-$e^{-1}) \sim$ 63$\%$ of the disc mass at $t=0$). The viscous time-scale can be then written as $t_\nu= R_{\rm c}^2/(3 \nu_{\rm c})$, where $\nu_{\rm c}=\nu(R_{\rm c})$. For our models, we thus have that $t_\nu\propto R_{\rm c}$

We numerically solve the gas viscous evolution equation 
\begin{equation}
\frac{\partial \Sigma_{\rm g}(R)}{\partial t} = - \frac{3}{R} \frac{\partial}{\partial R} \left( R^{1/2} \frac{\partial }{\partial R} ( \nu \Sigma_{\rm g}(R) R^{1/2}) \right),
\end{equation}
including also dust feedback as described in the next sub-section. As initial condition for the surface density of the gas $\Sigma_{\rm g}$, we use the analytical solution of \citet{LBP74}:
\begin{equation}
\Sigma_{\rm g}(R) = \frac{M_0}{2\pi R_{\rm c}^2}\left(\frac{R}{R_{\rm c}}\right)^{-1}\exp\left(-\frac{R}{R_{\rm c}}\right),
\label{eq:LB}
\end{equation}
where $M_0 = 2 \pi \int \Sigma R \mbox{d}R$ is the initial disc mass.
However, because we do not include substructures, the impact of dust feedback on the gas is very limited, thus the analytical self-similar solution of \citet{LBP74} is a very good description of the gas surface density at each time.

\subsection{Dust evolution}\label{sec:init_dust}

A fundamental parameter controlling the dust dynamics is the Stokes number $St$, defined as (in the Epstein regime)
\begin{equation}
St = \frac{\pi}{2}\frac{a \rho_{\rm dust}}{\Sigma_{\rm g}},
\end{equation}
where $a$ is the size of a dust grain, $\rho_d$ its bulk density and $\Sigma_g$ is the gas surface density. Grains with $St >> 1$ are decoupled from the gas, grains with $St<< 1$ are well coupled with the gas and grains with $S_t \sim 1$ experience the strongest radial drift.

The mass fraction of the dust components are evaluated as in \citet{Birnstiel2012}. Dust radially drifts and fragmentation is also included in the model. We use the radial drift limit or the fragmentation limit, whichever is lower, to set the maximum grain size\footnote{Hereafter only "grain size" for simplicity} $a_{\rm max}$ at each radius. If $a_{\rm max}$ is set by the fragmentation limit it corresponds to the maximum size that allows grains to collide without fragmenting, while in the radial drift case dust grains radially drift as fast as they grow. The two most important parameters to determine which regime is relevant are the value of the \citet{SS74} viscosity $\alpha$ and the fragmentation velocity $u_{\rm f}$, fixed in this work to 10 m/s. More details on the dust implementation in our model can be found in \citetalias{Rosotti2019} and \citet{Birnstiel2012}.

The dust radial drift velocity is computed according to the one fluid approach of \citet{Laibe2014}: it considers both the drag force due to the gas on the dust and the feedback of dust onto gas -- even if in our case the fast radial drift quickly decreases the amount of dust, making the feedback effect not significant.

\subsection{Initial conditions}\label{sec:init_cond}

The values of both $R_{\rm c}$ and $M_0$ affect the initial observed extent of the disc; viscous spreading affects its subsequent evolution and it depends on the value of $\alpha$. Thus, to test these dependencies, we set up a suite of models varying these parameters values in the typical range of variations found in the literature. 

For simplicity we fixed the stellar mass $M_\star=$1~M$_\odot$ and the initial gas to dust ratio to 100.
We selected three values of the viscosity $\alpha=10^{-2}, 10^{-3}, 10^{-4}$, that encompass the possible range of variation of this parameter. The highest value may be too high for the bulk of the disc population \citepalias[e.g.,][]{Trapman2020}, but still possible for individual discs \citep{Flaherty2020}. Indeed, even if $\alpha = 0.01$ used to be a common assumption \citep{WilliamsCieza2011}, modern studies suggest that weak turbulence ($\alpha \lesssim 10^{-3}$) may be common in protoplanetary discs \citep{Fedele2018,Long2018,Flaherty2020,Rosotti2020}. We set a lowest value of $\alpha$ as, working in the viscous framework, we need a mechanism able to explain the observational fact that discs are accreting; theoretically, this value could be given by pure hydrodynamical instabilities. The initial disc mass is important for the observed disc extent, both for the gas (because it determines where CO is able to self-shield against photo-dissociation, e.g., \citetalias{Trapman2019}) and the dust component \citepalias[because it determines the coupling of dust and gas and the extent of the large grains, those emitting in the sub-mm,][]{Rosotti2019}. Since measuring the disc mass is observationally challenging \citep[e.g.,][]{Williams2014,Miotello2017,BerginWilliams2017}, we then explored a plausible range of values $M_0 = 0.01, 0.03, 0.1$M$_\odot$. The initial disc size is also a poorly constrained value, due to the lack of high resolution observations of early stages of protoplanetary discs and to the difficulty to properly observe the size. Recent results from surveys such as CALYPSO \citep{Maury2019} or VANDAM \citep{Tobin2020}, as well as previous studies on the CO radii \citepalias{Trapman2020}, appear to suggest that class 0 and class 1 discs are compact. We therefore chose to test a conservative range of values: $R_0 = 10, 30, 80$ au.

\subsection{CO emission calculation}

To produce synthetic observations of the CO emission, we post-process all the time-steps of each model with the line radiative transfer code RADMC 3D \citep{Dullemond2012}. For simplicity, we assume that all our models are face-on discs. As discussed in \citetalias{Rosotti2019}, this assumption does not introduce any significant difference for the dust radii (as long as deprojection is correctly taken into account in the observations). This is in principle not true for line emission. Indeed, in an inclined disc the velocity component along the line of sight does not vanish: this modifies the shape of the emission lines. In addition, it increases the optical depth along the line of sight. \citetalias{Trapman2019} tested also this effect, finding that it can be neglected for inclinations smaller than $\sim$60\degr. It should be noted that deprojection correctly recovers the value of the radius only if we assume that the inclination of the disc is perfectly known; in reality, poorly constrained inclinations, which is the case at low signal to noise, can be a dominant source of error on radii estimates in observations \citep{Ansdell2018}. We plan to take this into account in the future by performing a population synthesis study, rather than the parameter study we undertake in this paper.

Starting from the gas and dust 1D surface density profiles, we interpolate the values of $\Sigma$ on a spherical grid with $r$ varying in $N_r=220$ steps from 1 to 5000~au and $\theta$ varying in $N_\theta = 80$ steps from 0 to $\pi$, building an azimuthally-symmetric disc by prescribing the vertical structure of the density $\rho$ as a Gaussian:
\begin{equation}
\rho(z) = \frac{\Sigma}{H\sqrt{2\pi}}\exp\left(-\frac{z^2}{2H^2}\right).
\end{equation}
We set the CO to H$_2$ abundance to $10^{-4}$ (but see below for regions of the disc where this does not apply) and we assume that CO molecules are in local thermal equilibrium (LTE). The molecular properties for CO are prescribed from the Leiden Atomic and Molecular Database \citep{Sholer2005}.  We include also a turbulent velocity dispersion of $v = 0.2$ kms$^{-1}$ to consider the small-scale broadening of the spectral lines due to chaotic motion.

To compute the molecular abundance we considered photodissociation and freeze-out of the CO molecules. Regarding photodissociation, we take into account self-shielding of CO and assume that the CO abundance vanishes whenever the column density in the vertical direction drops below the threshold value of $n_\mathrm{CO,th} = 1.672 \times 10^{-4}$ g cm$^{-2}$. This is a very simplified approach but it reproduces well results of more complex calculations \citep[e.g.,][]{vanDishoeckBlack1988,Facchini2016,Trapman2019}. Note that, due to the disc geometry, photo-dissociation along the vertical direction is always more important than photo-dissociation along the radial direction. To model freeze-out, we decrease the molecular abundance by a factor $10^{-3}$ when the temperature is below the 19 K threshold. 

The radiation field of the central star is modelled with a blackbody
spectrum, assuming $M_\star=1$M$_\odot$, $T_{\rm eff} = 4000$~K, $R_\star=2R_\odot$. 
We set the temperature in the midplane, $T_\mathrm{m}$, equal to the temperature used in the disc evolution calculation. For CO emission it is important to take into account the vertical temperature gradient and therefore we set the temperature in the disc upper layers $T_\mathrm{s}$ assuming optically thin heating and cooling \citep{Chiang1997,Dullemond2001}. The two temperatures are connected smoothly using the prescription first introduced by \citet{Dartois2003} and then modified by \citep{Rosenfeld2013}:
\begin{equation}
     T(r, z) = \left\{
\begin{array}{ll}
T_s + \left(T_m - T_s\right) \left[\sin \left(\frac{\pi z}{2 z_q}\right)\right]^{4} & \mbox{if $z < z_q$} \\
T_s & \mbox{if $z \ge z_q$} 
\end{array},
\right. 
\end{equation}
where we set $z_q=3H$. 
Once we have set up the model we compute the CO J=2-1 emission line, assuming Keplerian rotation, using 40 channels extending in radial velocity for $10$ km s$^{-1}$. Finally, from the resulting cube we compute the moment 0 (integrated intensity) map of each timestep of each disc.

To obtain the continuum 850 $\mu$m surface brightness of our models we compute the emission of the disc as in \citetalias{Rosotti2019} and \citetalias{Rosotti2019L}:
\begin{equation}
    S_b(R)=B_\nu(T(R))[1-\exp(-\kappa_\nu \Sigma_{\rm dust})], 
\end{equation}
where $B_\nu$ is the Plank function, $\kappa_\nu$ is the dust opacity and $\Sigma_{\rm dust}$ is the dust surface density.

 The opacity is calculated as in \citet{Tazzari2016}, using Mie theory and considering all the grains compact and spherical, composed by a misture of 10$\%$ silicates, 30$\%$ refractory organics and 60$\%$ water ice. To convert from the maximum grain size computed by the code at each radius to an opacity, we assume a power-law grain size distribution $n(a) \propto a^{-3.5}$ at each radius, with $a_{\rm min}<a<a_{\rm max}$.
\begin{figure*}
	\centering
	\includegraphics[width=0.9\textwidth]{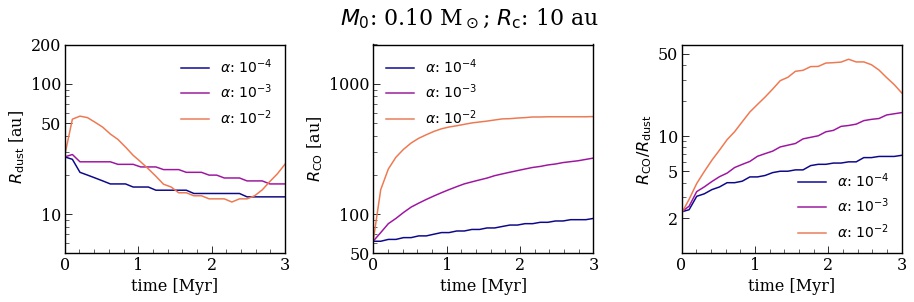}
    \caption{Dependence of the models on the value of the viscosity $\alpha$: right, middle and left panels shows the values of $R_{\rm dust}$, $R_{\rm CO}$ and the ratio $R_{\rm CO}/R_{\rm dust}$ as a function of disc age (in Myr) respectively, for different values of the viscosity $\alpha$ (blue: $\alpha=10^{-4}$, purple: $\alpha=10^{-3}$, orange: $\alpha=10^{-2}$) and discs with R$_0 = 10$~au and $M_0 = 0.1$~M$_\odot$. The behaviour of the dust radius has been already discussed in \citetalias{Rosotti2019}. Values of $R_{\rm CO}$ are in agreement with the findings of \citetalias{Trapman2020}. Note that, even for the lower values of viscosity, the values of the ratio $R_{\rm CO}/R_{\rm dust}$ are always larger than 4 for discs older than 1 Myr. }
    \label{fig:fig_1}
\end{figure*}

\subsection{Radius determination}

Our discs models are continuous structures where it is not straightforward to define a radius. We decided to define the observed dust and gas radii for the two tracers, $R_{\rm CO}$ and $R_{\rm dust}$, as the radius that contains 68$\%$ of the flux of the given tracer. The value of the fraction of the flux to use is arbitrary; we choose these values as they have been used by previous observational studies, e.g., \citet{Sanchis2021} for $R_{\rm gas}$ and \citet{Thripati2017} and \citet{Tazzari2020} for $R_{\rm dust}$. Moreover, considering the 68$\%$ flux radius rather than a higher fraction (e.g., 90$\%$) reduces the observational uncertainties due to lower signal to noise values for faint sources.

While to perform quantitative comparison with observations it is necessary to compute these radii from the numerical models, it is helpful to keep in mind that these radii have qualitative physical interpretations that can help us to interpret the results we will present in the following sections. For what concerns the gas radius, through extensive thermochemical modelling \citetalias{Trapman2019} showed that this radius closely tracks the radius at which CO is photo-dissociated when its column density becomes too low to self-shield. Because CO is very resistant to dissociation due to self-shielding, in practice this always happens at a radius that encloses a large fraction of the disc mass; therefore, the observed $R_{\rm gas}$ inherits this propriety and most of the disc mass is contained within $R_{\rm gas}$. Note that this implies a strong dependence on the shape of the surface density in the outer part of the disc, in our model the exponential tail of the self-similar profile. \citetalias{Trapman2019} also showed that this radius is independent from the dust evolution. Here, we consider the J=2-1 CO emission line but we tested also the J=3-2 CO emission line case, finding no significant differences in the values of the radii (the difference is below 2$\%$).
For what concerns the dust radius instead, as shown in \citetalias{Rosotti2019} and \citetalias{Rosotti2019L}, the 68$\%$ $R_{\rm dust}$ at a given wavelength is related to the position where the maximum grain size is comparable to the wavelength:
because of the presence of an \textit{opacity cliff} (a resonance in the opacity when grain size and wavelength are comparable), the opacity drops quickly for grain populations with $a<a_{\rm max}$. Since the maximum grain size is set by either fragmentation or radial drift, these two processes, and the parameters controlling them (such as the level of turbulence, the dust-to-gas ratio, the fragmentation velocity) set the observed dust radius.

Operationally, to estimate the disc radii we apply the \textit{cumulative flux} method, calculating for each timestep of each model the flux on increasingly larger concentric rings centred on the source. Once the cumulative flux is computed, we select the radius that has the threshold value.

\section{Results}\label{Results}
\subsection{Disc radius evolution: dependence on disc viscosity}

To better illustrate our results, we first describe in detail an example, fixing the disc parameters $R_{\rm c}$ and $M_0$ and varying the value of the viscosity $\alpha$. After this we will show how the results change when changing different parameters of the disc.

Figure \ref{fig:fig_1} shows in the top, middle and bottom panels the values of $R_{\rm dust}$, $R_{\rm CO}$ and the ratio $R_{\rm CO}/R_{\rm dust}$, respectively, as a function of disc age (in Myr), for different values of the viscosity $\alpha$. The cases we show have a disc initial mass of $M_0=0.1$ M$_\odot$ and an initial characteristic scale of $R_{\rm c}=10$ au, in agreement with the fact that observation in Lupus are consistent with viscous evolution if discs are initially small ($\sim$10 au) \citepalias{Trapman2020}. The value of $\alpha$, for a fixed $R_{\rm c}$, affects the initial viscous timescale of the discs, which spans from 0.05 to 5 Myr.
\begin{figure}
	\centering
	\includegraphics[width=0.9\linewidth]{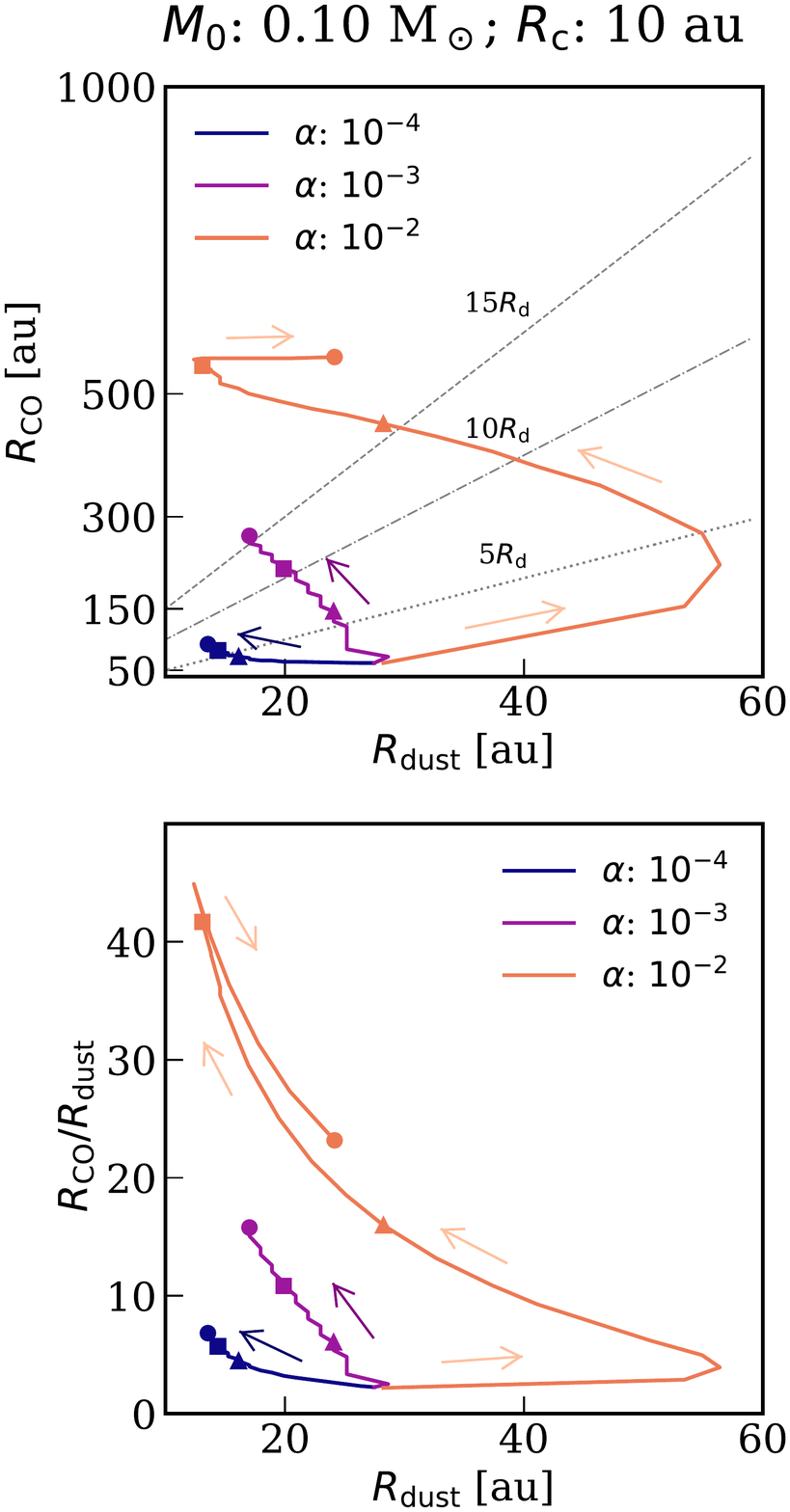}
    \caption{Tracks for $R_{\rm CO}$ (top) and the ratio $R_{\rm CO}/R_{\rm dust}$ (bottom) as a function of $R_{\rm dust}$ respectively, for different values of the viscosity $\alpha$ (blue: $\alpha=10^{-4}$, purple: $\alpha=10^{-3}$, orange: $\alpha=10^{-2}$) and discs with R$_0 = 10$~au and $M_0 = 0.1$~M$_\odot$. Values of all the quantities at 1,2 and 3~Myr are shown with triangles, squares and circles respectively. To help the reader, the flow is indicated by colourful arrows. The top figure also display with dotted, dash-dotted and dashed lines $R_{\rm CO}=5R_{\rm dust}, 10R_{\rm dust}, 15R_{\rm dust}$.}
    \label{fig:fig_2}
\end{figure}

The behaviour of $R_{\rm dust}$ (left panel) is discussed in detail in \citetalias{Rosotti2019}, here we just want to underline that for small and intermediate values of $\alpha$ ($10^{-4}$ and $10^{-3}$) viscous spreading happens at a slow rate and the dust radii decrease with time; the observed dust radii are relatively small, $R_{\rm dust}$ $\lesssim 20-30$~au. This is a consequence of radial drift depleting the disc of large grains; because of the opacity cliff only grains large enough (that is, with a size comparable to the wavelength) have significant opacity to be observed and therefore the disc size decreases with time. For large values of $\alpha$ ($10^{-2}$) there are two phases of expansion at different rates: a very rapid initial one due to viscous spreading while the grains are still growing and well coupled with the gas, and a second one, after 2 Myr, where the value of the observed dust radius increases with time: this is because after this point all the dust grains are below the opacity cliff\footnote{The opacity cliff will affect also the models with smaller values of $\alpha$, increasing the values of $R_{\rm dust}$, on timescales longer than the ones considered}. However, note that this second expansion is not due to spreading: what has happened is that the  bright disc emission close to the star (where the large grains are) has disappeared due to radial drift and now only a faint \textit{halo} of small grains is observable.

The behaviour of $R_{\rm CO}$ (shown in the middle panel of Fig.~\ref{fig:fig_1}) is obtained with a post-processing of the models using the RADMC code, considering only prescribed CO chemistry. However, our findings are in agreement with what \citetalias{Trapman2020} found using the thermochemical code DALI, that consider a more complex chemistry for the disc (for a comparison between the models see Appendix~\ref{App_1} and Fig.~\ref{fig:A_1}). The value of $R_{\rm CO}$ is much larger (in this case, initially it is 6 times larger, but can be up to 10 times larger according to \citetalias{Trapman2020}) than the value of $R_{\rm c}$ for all the values of the viscosity and all the times. Indeed, the surface density threshold for the CO photodissociation is very low, and it is reached for values much larger than $R_{\rm c}$. Moreover, where CO is not photodissociated, it is optically thick, thus a small mass fraction can significantly contribute to the flux. For an extensive discussion on this topic see \citetalias{Trapman2019}.

Fig.~\ref{fig:fig_1} (middle panel) shows how, depending on the value of the viscosity, the gas radii grow with time. Two opposite effects are at play: viscous spreading, resulting in an increase of the radius, and the enhanced photodissociation of CO in the outer part of the disc as the surface density decreases, resulting in smaller values of the gas radius. Viscous spreading is dominant between the two effects for most of the parameter space \citepalias{Trapman2020} and the gas radius of the disc $R_{\rm CO}$ increases with time, reaching after 3~Myr the value of $\sim 70-200$~au for small and moderate viscosities. For higher values of $\alpha$, viscous spreading is substantial in the first Myr, and $R_{\rm CO}$ has increased by a factor 2-3 after 3 Myr.

Fig.\ref{fig:fig_1} (right panel) displays the ratio $R_{\rm CO}/R_{\rm dust}$ as a function of the disc age. The quantity already starts with a relatively high value, 2, and monotonically increases for low and moderate values of $\alpha$, reaching values $\sim 10$ after 3 Myr, while for high values of $\alpha$ it grows until $\sim 50$ before decreasing, after 2 Myr, to $\sim 20$ thanks to the second phase of increase of $R_{\rm dust}$. In summary, even for very low values of $\alpha$, this ratio is larger than 4 for discs older than 1 Myr.
\begin{figure*}
	\centering
	\includegraphics[width=0.9\textwidth]{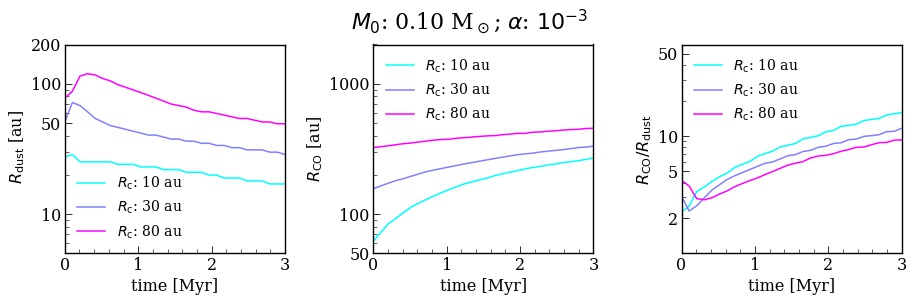}
    \caption{Dependence of the models on the value of the initial scale radius $R_{\rm c}$: left, middle and right panels shows the values of $R_{\rm dust}$, $R_{\rm CO}$ and the ratio $R_{\rm CO}/R_{\rm dust}$ as a function of disc age (in Myr) respectively, for different values of the initial radius (cyan: $R_{\rm c}$=10~au, violet:  $R_{\rm c}$=30~au, magenta:  $R_{\rm c}$=80~au) for discs with $\alpha = 10^{-3}$ and $M_0 = 0.1$~M$_\odot$. For all the quantities, the value of $R_{\rm c}$ is the less effective, affecting the discs as a scale factor.}
    \label{fig:fig_3}
\end{figure*}

The evolutionary tracks for $R_{\rm CO}$ and the ratio $R_{\rm CO}/R_{\rm dust}$ as a function of $R_{\rm dust}$ for the models are shown in Figure \ref{fig:fig_2}, top and bottom respectively. The values of the quantities for 1, 2 and 3~Myr are represented with triangles, squares and dots respectively, and to guide the reader we also draw the lines where $R_{\rm CO}=5, 10$ and 15 $R_{\rm dust}$. As the plots show, after $\sim$ 1~Myr the value of  $R_{\rm CO}$ is at least 4 times larger than $R_{\rm dust}$ for all the tested values of viscosity, and remains $\sim$ 10$_{\rm dust}$ for a few Myr for small and moderate values of viscosity, reaching $\sim$ 15$R_{\rm dust}$ even in case of moderate $\alpha$. For $\alpha=10^{-2}$ the ratio reaches very large values after $\sim$ 1 Myr, larger than 40, to then decrease after 2~Myr to $\sim$ 20. From these plots is evident that, for most of the tested values of the age of the discs, the value of the ratio $R_{\rm CO}/R_{\rm dust}$ we should expect to observe should be $\gtrsim 5$. 

\subsection{Dependence on the initial scale radius}
\begin{figure*}
	\centering
	\includegraphics[width=0.9\textwidth]{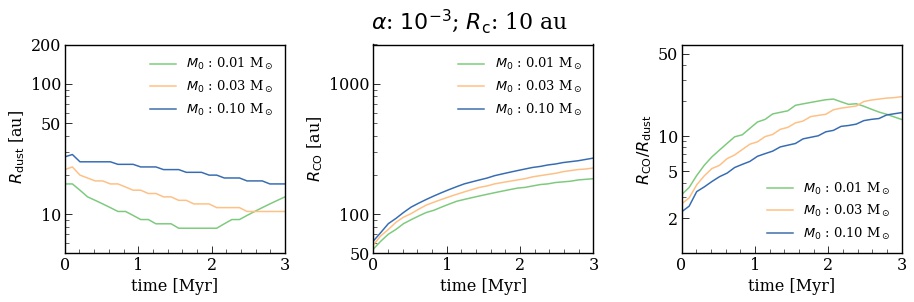}
    \caption{Dependence of the models on the value of the initial disc mass $M_0$: left, middle and right panels shows the values of $R_{\rm dust}$, $R_{\rm CO}$ and the ratio $R_{\rm CO}/R_{\rm dust}$ as a function of disc age (in Myr) respectively, for different values of the disc mass (green: $M_0$=0.01~M$_\odot$, yellow: $M_0$=0.03~M$_\odot$, blue:  $M_0$=0.1~M$_\odot$) for discs with $\alpha = 10^{-3}$ and $R_{\rm c}$ = 10~au. Initially lighter discs ($M_0$=0.01~M$_\odot$) experience also the second expansion due to the opacity cliff effect.}
    \label{fig:fig_4}
\end{figure*}

In this section we explore the effect of changing $R_{\rm c}$, keeping the disc mass fixed at $10^{-1}$M$_\odot$ and an intermediate value for $\alpha =10^{-3}$, an intermediate value in the range admitted by MRI (see Sec.~\ref{sec:init_cond}).  

In Fig.~\ref{fig:fig_3} we show our results for models obtained varying $R_{\rm c}=10, 30, 80$ au. Changing the value of $R_{\rm c}$ affects the viscous time scale of the discs, which spans from 0.5 to 3 Myr.
The left, middle and right panels show the values of $R_{\rm dust}$, $R_{\rm CO}$ and the ratio $R_{\rm CO}/R_{\rm dust}$ as a function of disc age (in Myr), respectively. 
As expected (see also \citetalias{Rosotti2019}),  the dust flux radius shrinks with time in all models (Fig. ~\ref{fig:fig_3}, left). After 3~Myr the final size of the dust radius is larger for larger $R_{\rm c}$ and decreases with the same slope for all the models, related to the value of $\alpha$. Initially larger discs ($R_{\rm c}= 30,80$ au) experience also with moderate values of the viscosity the very rapid initial expansion in the dust we described in the previous section. We recall that this phase lasts the time the grains need to grow from the initial, sub-$\mu$m sizes up to the limit imposed by radial drift (a few $10^5$ yr, longer for larger $R_{\rm c}$).  

Fig.~\ref{fig:fig_3} (middle) show
the value of the observed gas radii for the models. Viscous expansion is initially faster for smaller discs (the discs are $\sim$ 2.5 and 1.4 times larger after 1 Myr and $\sim$ 3.5 and 1.7 times larger after 3 Myr, respectively), while the effect of viscous spreading is reduced for initially larger discs (the disc size increases only by a factor 1.3 after 3 Myr for the R$_0 = 80$ au case). Thus, while the disc radii of initially smaller discs grow due to viscous expansions in the first Myr, for larger discs the viscous timescale is long enough that the evolution in the first 3 Myr is minor. For a detailed discussion on the observational implication of the allowed values of viscosity see \citetalias{Trapman2020}. 
After 3~Myr initially small discs have grown to relatively large sizes, about $\sim 300$ au, and initially larger disc can reach $\sim 400$ au.

The result of the combined evolution of these quantities is shown in Fig.~\ref{fig:fig_3} (right): in the initial phase ($t\lesssim 5 \times 10^{5}$ yr) the ratio $R_{\rm CO}/R_{\rm dust}$  decreases down to 2-3 for initially larger discs, then increases with time reaching high values\footnote{Compared with the typical observed values for disc populations, $\sim 2-4$. However, extreme discs show values $\gtrsim 5$ \citep[e.g.,][]{Facchini2019,Sanchis2021,Kurtovic2021}} ( > 5), even if the value of the viscosity is moderate ($\alpha=10^{-3}$). Note also that initially larger discs have smaller values of this ratio compared with initially smaller disc. This is due to the combination of the two effects we described previously: the initial expansion of the dust radius and the smaller increase of the gas radius.

We can conclude that the initial characteristic scale $R_{\rm c}$ affects the models as a scale factor for the size. While it introduces a small trend for the gas/dust ratio $R_{\rm CO}/R_{\rm dust}$ (larger discs have a smaller ratio), it does not have a critical influence on the overall values, which are always large (>5). 

\subsection{Dependence on the initial disc mass}

Fig.~\ref{fig:fig_4} displays our results for models obtained varying the initial disc mass $M_0=0.01, 0.03, 0.1$~M$_\odot$. In this case, where $R_{\rm c}=10$~au and $\alpha=10^{-3}$, the viscous timescale is fixed to 0.5 Myr.
The left, middle and right panels show the values of $R_{\rm dust}$, $R_{\rm CO}$ and the ratio $R_{\rm CO}/R_{\rm dust}$ as a function of disc age (in Myr) respectively. Looking at Fig.~\ref{fig:fig_4} (left) we can follow the time evolution of $R_{\rm dust}$ for different values of the disc mass. After a very quick transition phase, all the models are in the drift dominated regime, thus the value of the observed dust disc size decreases with time. Initially lighter discs ($M_0=0.01$~M$_\odot$) experience also the second expansion at 2~Myr when all the dust grains become too small and move below the opacity cliff. The reason why this happens at earlier times is simply because these discs, due to the smaller initial mass, run out of large grains at earlier times. We expect to have this effect also for the other cases on longer timescales, not shown on the plot. 
In the middle panel of Fig.~\ref{fig:fig_4} we show the time evolution of $R_{\rm CO}$ for the three models. Again, we find that the gas disc size  increases with time, and that more massive discs are slightly larger, reaching the value of $\sim 200$~au after 3 Myr. The time evolution of $R_{\rm CO}/R_{\rm dust}$ is shown in the right panel of Fig.~\ref{fig:fig_4}: the effect of larger $M_0$ is to have a slightly smaller ratio. The lightest disc show a clear decrease in the ratio after 2~Myr when the dust size increases. However, in all the models the value of the ratio, initially $< 5$, becomes larger than 10 after 2 Myr.

In summary, the initial disc mass $M_0$ affects the models as a scale factor for the surface density. The mass influences both the size of the dust grains (and consequently the effect of the opacity cliff) and CO photodissociation (since this becomes effective when the surface density is below a critical value). However, the effect of varying the mass is relatively modest and does not affect significantly the resulting values of the ratio $R_{\rm CO}/R_{\rm dust}$.

\begin{figure*}
	\centering
	\includegraphics[width=0.85\textwidth]{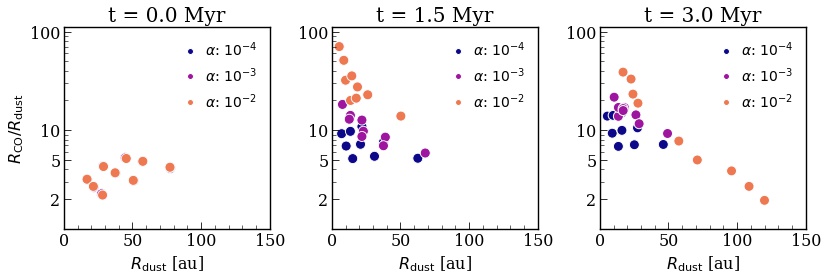}
    \caption{Three different snapshots of $R_{\rm CO}/R_{\rm dust}$ vs $R_{\rm dust}$ (in au) at 0 (left),1.5 (middle) and 3 (right) Myr respectively. Different values of viscosity are shown with different colors ($\alpha=10^{-4}$ blue, $\alpha=10^{-3}$ purple, $\alpha=10^{-2}$ orange); Initially (0 Myr) the points overlap, hence we only see $\alpha = 10^{-2}$ in the figure. Note that the value of $R_{\rm CO}/R_{\rm dust}$ for all the discs is always larger than 6 at 1.5~Myr and $R_{\rm dust}$ is smaller than $\sim$ 50 au. After 3 Myr, only the discs with high viscosity have $R_{\rm CO}/R_{\rm dust}$ $\lesssim$ 5, with very large values of $R_{\rm dust}$ ( > 60 au).}
    \label{fig:fig_6}
\end{figure*}

\subsection{Observable quantities from the models}\label{sec:obs}

In the previous sections we have described the evolution of the disc sizes with time. Since it is not possible to observe this evolution due to the long timescales involved, we now ask the question of how our models would appear if observed at a given time. We show in Fig.~\ref{fig:fig_6} three different scatter plots representing snapshots of $R_{\rm CO}/R_{\rm dust}$ vs $R_{\rm dust}$ (in au) at 0 (left),1.5 (middle) and 3 (right) Myr respectively, underlining the different values of viscosity with different colours. A more detailed version of Fig.~\ref{fig:fig_6}, that identifies all the discs with size, mass and viscosity value, is shown in Appendix~\ref{App_2}. The left plot show the initial conditions ($t=0$~Myr) of the models, which consist in a simple parameter space exploration. We will expand our study in the future by performing population synthesis, which will also allow us to introduce an age spread among the discs. Note how initially our discs have $R_{\rm dust}$ smaller than 90~au and $R_{\rm CO}/R_{\rm dust}$ smaller than 6; visually, only the bottom-left part of the plot is populated. Covering a larger area would require either increasing the initial disc mass or the disc radius. Neither option is viable since we already consider discs marginally gravitationally unstable and \citetalias{Trapman2020} showed that initially larger discs, after viscous spreading, would be too large in comparison with observations.

The middle plot of Fig.~\ref{fig:fig_6} shows the values of $R_{\rm CO}/R_{\rm dust}$ vs $R_{\rm dust}$ after 1.5~Myr. Due to the effect of radial drift, the dust radius $R_{\rm dust}$ is smaller and the gas radius $R_{\rm CO}$ is larger than the initial value: the combined effect is that  $R_{\rm CO}/R_{\rm dust}$ is > than 4 for all the discs. Visually, only the top - left part of the plot is populated, meaning that surveys of very young (1.5~Myr) discs are expected to measure $R_{\rm CO}\gtrsim 5-10$ or even $\gtrsim 10$ times larger than $R_{\rm dust}$ depending on the value of the viscosity. Note that for higher values of the viscosity, $R_{\rm CO}/R_{\rm dust}$ can be also larger than 20, with $R_{\rm dust}$ values smaller than 30~au. For low and intermediate values of viscosity the dust disc size $R_{\rm dust}$ of the models can be larger (> 30 au, with a few discs\footnote{Initially larger in size, $R_{\rm c}=$80~au} with size $\sim$ 60 au).

In the right plot of Fig.~\ref{fig:fig_6} we show the values of $R_{\rm CO}/R_{\rm dust}$ vs $R_{\rm dust}$ after 3~Myr. While after 1.5~Myr all the points concentrate in the top left part of the plot, here some models have moved to the bottom right, meaning that they have small $R_{\rm CO}/R_{\rm dust}$ (< 5) values but large dust radii (> 80~au). All these models have high values of viscosity ($\alpha=10^{-2}$) and small initial disc mass ($M_0=0.01~$M$_\odot$\footnote{See Fig.~\ref{fig:A_3} and Fig.~\ref{fig:A_4}}): these are discs that underwent the second expansion of the dust radius after 2~Myr. This expansion, which happens when all the grains become smaller than the opacity cliff, increases the apparent value of $R_{\rm dust}$, moving the points towards the right part of the plot. The gas radius of the disc, governed by the combined effect of the viscous spreading and photodissociation, remains relatively small (or starts to decrease after 1 Myr). The $R_{\rm CO}/R_{\rm dust}$ value is then smaller for these discs. All the other models remain very compact in the dust radii, with $R_{\rm dust}$ smaller than 40~au, but still have very large values of $R_{\rm CO}/R_{\rm dust}$ (> 6). Again, the higher the value of the viscosity, the higher the value of $R_{\rm CO}/R_{\rm dust}$. In summary, from this plot we would expect to observe in young star forming regions a broad sample of sources with large values of $R_{\rm CO}$ and small values of $R_{\rm dust}$ and a few sources with very large dust radii and very compact gas radii.

\section{Discussion}\label{Discussion}

\subsection{Comparison with observations}

\begin{figure}
	\includegraphics[width=0.8\columnwidth]{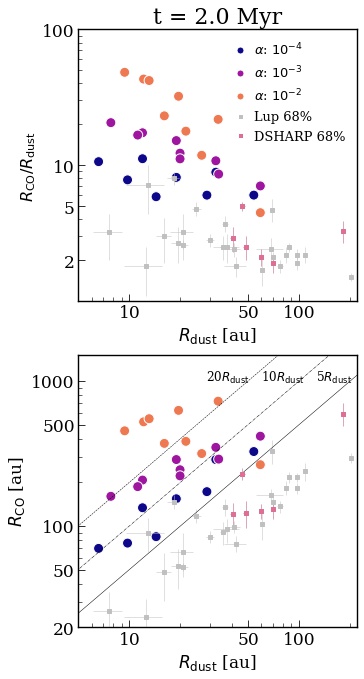}
    \caption{Comparison between the models at 2~Myr (dots) and the observed values in Lupus (squares) of $R_{\rm CO}/R_{\rm dust}$ (top) and $R_{\rm CO}$ (in au, bottom) vs $R_{\rm dust}$ (in au). Different values of viscosity are shown with different colours ($\alpha=10^{-4}$ blue, $\alpha=10^{-3}$ purple, $\alpha=10^{-2}$ orange). $R_{\rm CO}$ and $R_{\rm dust}$ values in Lupus shown in light grey are from \citet{Sanchis2021} and \citet{Hendler2020}, the pink squares are the discs observed also in the DSHARP survey \citep{Andrews2018}. In the bottom figure are also shown the lines $R_{\rm CO}$= 5,10,15 $R_{\rm dust}$ with solid, dashed and dotted lines respectively.} 
    \label{fig:fig_7}
\end{figure}

To test population synthesis models we need samples of discs that are homogeneous in age, in the environment and in the methodology of measurements, i.e. observations where both gas and dust are spatially resolved and $R_{\rm dust}$ and $R_{\rm CO}$ are evaluated with the same criteria (in our case the 68$\%$ gas and dust flux values obtained with the cumulative flux method). 
Natural candidates are the results coming from surveys of young ($\sim$2 Myr) star forming region such as Lupus. 
The comparison between the data and the models is shown in Fig.~\ref{fig:fig_7}. In the top panel we overplot the $R_{\rm CO}/R_{\rm dust}$ values obtained with our 2~Myr models and the observed $R_{\rm CO}/R_{\rm dust}$ values. 
For a more complete analysis we also show a comparison between the observed values of $R_{\rm CO}$ vs $R_{\rm dust}$ and our values for models at 2~Myr (Fig.~\ref{fig:fig_7}, bottom). 

The Lupus sample has been selected from the observed targets of \citet{Sanchis2021}, where gas and dust (68$\%$) radii have been measured in 42 sources, excluding 10 sources where only upper limits were detected; we added another source (SZ 98 / HK Lup) observed in gas by \citet{Ansdell2018} and in dust by \citet{Hendler2020}. Discs observed also in DSHARP and included in \citet{Sanchis2021} are: Sz 68 (HT Lup), Sz 71 (GW Lup), Sz 82 (IM Lup), Sz 83 (RU Lup), Sz 114, Sz 129 and MY Lup. All of them have resolved substructures. Moreover, J1608 (J16083070-3828268) has been recently imaged with ALMA and SPHERE as a disc with two bright lobes and a large cavity \citep{Marion2019}, Sz 98 / HK Lup is not well reproduced with smooth models \citep{Tazzari2020} and Sz 91 has been resolved in an axisymmetric ring \citep{Tsukagoshi2019}.

Discs in Lupus have dust radii $R_{\rm dust}$ that span from 20 to 80~au, with a few very wide discs with  $R_{\rm dust}\gtrsim 100$.
On the contrary, all the models with low and moderate values of the viscosity ($\alpha=10^{-4}, 10^{-3}$) are located in the top-left part of the plot, having compact dust radii and very large gas radii, $R_{\rm CO}$ $\sim 50-400$~au (see Fig.~\ref{fig:fig_7} bottom panel), evidencing dust evolution and an efficient radial drift. These models match a few points from the Lupus survey, where $R_{\rm CO}/R_{\rm dust}$ > 4 and  radial drift is expected to play a key role \citep{Facchini2019}. All these discs are discussed in \citealt{Sanchis2021} and are either wide binaries (Sz 75/ GQ Lup, Sz 65), faint discs (Sz 131) or very active discs with cloud contamination (Sz 82/ IM Lup, Sz 83/ RU Lup). Only a few models with high value of $\alpha=10^{-2}$ (that we recall are the smaller and lighter, see Sec.~\ref{sec:obs}) or large initial disc size ($R_0=80$ au) have smaller $R_{\rm CO}/R_{\rm dust}$ and large dust radii, covering the bottom-right part of the plot close to where observed values of discs with substructures are. 

Looking at the gas radii, Fig.~\ref{fig:fig_7} (bottom), we see that almost all the observational points lie below the line $R_{\rm CO}= 5 R_{\rm dust}$, while all the models have values above the same line: the $R_{\rm CO}$ values obtained from the models are between 5 and 10 times $R_{\rm dust}$ in case of low viscosity ($\alpha=10^{-4}$) and between 10 and 20 times $R_{\rm dust}$ for $\alpha=10^{-3}$, thus reaching very large gas radii with relative compact dust radii (< 50~au). Even higher values can be reached for high values of the viscosity ($\alpha=10^{-2}$).
This plot shows how high values of viscosity or large initial disc sizes fail to reproduce the observed gas radii, in agreement with \citetalias{Trapman2020}. Note that the $R_{\rm CO}$ values obtained for low and moderate viscosity themselves are compatible with the observed gas sizes, but their associated dust radius is much smaller than observed.

The same figures obtained with the 90$\%$ dust and gas radii ($R_{\rm dust,90\%}$ and $R_{\rm CO,90\%}$) evaluated with the cumulative flux method are shown in Appendix \ref{App_3}. Considering a different threshold for the gas and dust radii changes their values for the single objects both for the observations and for the models, obtaining as expected larger values for higher thresholds of the flux. However, the $R_{\rm CO}/R_{\rm dust}$ values have no significant difference in the population with respect to the 68$\% R_{\rm dust}$ and $R_{\rm CO}$, still being between 2 and 4. 

Summarising, Fig.~\ref{fig:fig_7} clearly show that our models do not reproduce the observed dust and gas radii in surveys of star forming regions for smaller and moderate values of viscosity. Considering a more evolved population (t = 3Myr, see Fig. \ref{fig:fig_6}) would not change our analysis. Models with $\alpha=10^{-2}$ and initially small and compact discs or with $\alpha=10^{-3},10^{-4}$ and initially large size ($R_0 = 80$ au) seem to match some sources in Lupus (the DSHARP observed sample, J1608, HK Lup), known for being quite wide in dust flux, more massive and for having substructures \citep{Long2018}. However, as found by \citetalias{Trapman2020}, discs with $\alpha=10^{-2}$ have a too large $R_\mathrm{CO}$ compared to the bulk of the disc population. 

It should be kept in mind that we fixed the value of the mass of the star to 1~M$_\odot$ while many of the observed targets in surveys have stellar masses smaller than M$_\odot$ (see \citealt{Ansdell2018} or \citealt{Hendler2020} for Lupus).  
In any case, according to the results of Lupus survey of \citet{Sanchis2021}, the bulk of the disc population appears to have a similar behaviour and evolutionary stage, independent of the stellar and disc properties.
We defer to a future work a population synthesis study in which we Montecarlo sample the disc initial conditions. 

\subsection{Possible solutions}

In this section we discuss three possible ways to resolve the apparent discrepancy between our models and the observed data: (\emph{i}) a slower grain growth process; (\emph{ii}) the impact of disc-wind driven evolution and (\emph{iii}) the impact of substructures.

\subsubsection{Dust evolution: a slower growth?}\label{sec:DG}
\begin{figure}
	\centering
	\includegraphics[width=0.75\linewidth]{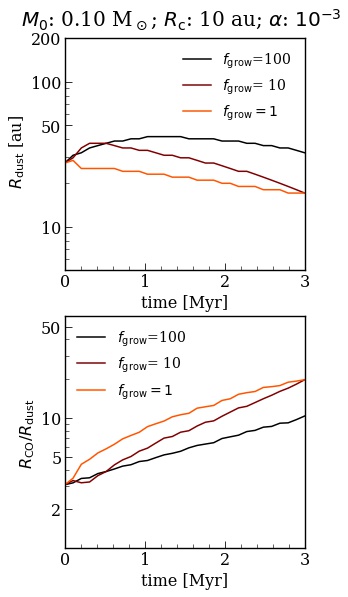}
    \caption{Dependence of the fiducial model ($M_0=0.1$~M$_\odot$, $\alpha = 10^{-3}$ and $R_{\rm c} = 10$~au) on the value of the growing factor $f_{\rm grow}$: top and bottom panels shows the values of $R_{\rm dust}$ (top) and the ratio $R_{\rm CO}/R_{\rm dust}$ (bottom) as a function of disc age (in Myr) respectively, for different values of the the dust growth timescale. The fiducial model with growing factor $f_{\rm grow}=1$ is shown in orange, models with $f_{\rm grow}=1/10$ and $f_{\rm grow}=1/100$ are shown in dark red and black respectively. }
    \label{fig:fig_8}
\end{figure}

A pivotal role in our model is played by the dust properties such as the grain size distribution (single or two populations), the initial size distribution (that impacts on when radial drift occurs) and the local maximum grain size $a_{\rm max}$ (crucial to determine the dynamics of the dust and the mm emission of the discs). In the two population model of \citealt{Birnstiel2012} the dust grains of size $a$ grow until they reach $a_{\rm max}$ (see Sec.~\ref{sec:init_dust}) on a timescale t$_g$ defined as
\begin{equation}
    t_g = \frac{a}{\dot{a}} = f_{\rm grow} \left( \frac{1}{\Omega} \frac{\Sigma_g}{\Sigma_d} \right),
\end{equation}
where $\Omega$ is the Keplerian angular frequency and the growing factor $f_{\rm grow}$ is an extra factor (considered to be equal to 1 in the "standard" two population model and in this work) used e.g in \citealt{Booth2020} to measure the fraction of collisions that lead to growth: $f_{\rm grow} =1$ means that all the collisions result in dust growth while $f_{\rm grow} =100$ means that only $1\%$ of the collisions result in growth, leading to longer lifetimes for dust in discs due to a less efficient radial drift (for an extended discussion see \citealt{Booth2020} or \citealt{Sellek2020}). 

To test whether prescribing a less efficient radial drift could solve the discrepancy between our models (see Fig.~\ref{fig:fig_7}) and observed data in Lupus, we decided to vary the fraction of collisions leading to grain growth $f_{\rm grow}$, testing also the cases $f_{\rm grow}=10$ and $f_{\rm grow}=100$. Indeed, results from \citealt{Booth2020} suggested that $f_{\rm grow} >1 $ could slow down the removal of large dust grains, thus increasing the observed dust radius $R_{\rm dust}$ after 3 Myr. 

Fig.~\ref{fig:fig_8} shows the results of varying $f_{\rm grow}$ on the fiducial model $M_0=0.1$~M$_\odot$, $\alpha = 10^{-3}$ and $R_{\rm c}$ = 10~au.
The "classic" model f$_{\rm grow}=1$ is shown in orange, while f$_{\rm grow}=10$ and f$_{\rm grow}=100$ are plotted with dark red and black lines. The top figure shows the time evolution of $R_{\rm dust}$ (in au) as a function of time (in Myr). The gas evolution is not dependent on the dust growth, thus the gas radii values are the same of our previous models (not shown in fig. \ref{fig:fig_8}). For the case $f_{\rm grow}=10$, after an initial difference in the dust radius for $t < 2$ Myr\footnote{that could be dependent on the initial conditions of the models and is difficult to test with observations, being the discs still embedded in their parental clouds}, the observed dust radius shrinks to $\sim$ 20~au at 3 Myr, compatible with the standard case, resulting in an almost equal values of the $R_{\rm CO}/R_{\rm dust}$ ratio (Fig. \ref{fig:fig_8}, bottom panel), too large with respect to the observed values. An extreme slow down of the timescale of radial drift ($f_{\rm grow}=100$, implying a timescale 100 times slower), could practically block the dust evolution, leading to a larger dust radius at all the timescales. Initially dust and gas are well coupled and the dust radius increases up to $\sim$ 40~au; after 2~Myr grains are still growing but radial drift starts to shrink the dust radius. This results in doubling $R_{\rm dust}$ at 3 Myr,  $R_{\rm dust} \sim$ 30~au, halving the  $R_{\rm CO}/R_{\rm dust}$ ratio. 

Generalising these results to all the initial conditions we tested, assuming an extremely ($f_{\rm grow}=100$) less effective grain growth (implying also a less efficient radial drift) could explain the observed larger values of $R_{\rm dust}$, lowering the $R_{\rm CO}/R_{\rm dust}$ values. However, this scenario seems unlikely. Indeed, in this case, the disc radius after 3~Myr is very similar to the initial value, implying a strong dependence on the initial conditions of protoplanetary discs and no disc evolution: initially large discs remains large while initially compact discs are observed with small dust radii. Moreover, \citealt{Sellek2020} found that the resulting dust mass of the discs are proportional to $f_{\rm grow}$, so that a large reduction in growth efficiency also corresponds to a correspondingly large increase in dust mass: that would help to explain the observed large masses of a few discs in Lupus, but the bulk of the population would be too massive compared to observations.

\subsubsection{Gas evolution: winds vs viscosity}\label{sec:DW}

Two mechanisms are currently invoked to explain the evolution of the gas in protoplanetary discs: viscous evolution and disc-wind models. 
In the first scenario \citep{LBP74}, implemented in our models, discs evolve for the angular momentum conservation and redistribution. 
In the latter model, magnetic disc winds remove angular momentum rather than transporting it through the disc (see e.g., \citealt{Suzuki2016}), thus the discs sizes should not grow with time. 
The timescales of disc evolution strongly depends on the strength of the magnetic field and the amount of magnetic flux inside the discs \citep{Bai2016}. 

Assuming that dust evolution is driven by pure radial drift, thus obtaining the same observed dust radius, if this second mechanism is at play the gas disc size is expected to be constant or, under some conditions \citep{Bai2016}, decrease with time. Initially larger discs ($R_0>30$ au) could lead to large dust radii and smaller gas radii, in agreement with the observed data. This would reduce the ratio $R_{\rm CO}/R_{\rm dust}$, obtaining values more consistent with observations. However, according to our models, the gas radius should be reduced by a factor 2-4: all the observations are below the line $R_{\rm CO}=5R_{\rm dust}$ while almost all the models have $R_{\rm CO} > 5 R_{\rm dust}$ (see bottom panel of Fig.~\ref{fig:fig_7}). This gives a very strong constraint on the efficiency of disc winds, that have to be able to remove a large fraction of the disc mass. Considering that the largest dust radius we obtain in our model is $\sim 50$ au, using $R_{\rm c}=80$ au, we would need to invoke that initially discs are roughly twice this size to reproduce the observed dust sizes.
At face value this does not seem compatible with the dust sizes observed in young class 0 and I discs \citep{Tobin2020}.
For these reasons, it seems unlikely that the observed $R_\mathrm{CO}/R_\mathrm{d}$ ratio could be explained invoking disc winds.

Other effects, such as photo-evaporation given by the high-energy X-ray and UV flux coming from the central star \citep{Clarke2001,Owen2012} or from nearby massive stars \citep{Facchini2016} contribute to the evolution of discs, shortening their lifetimes and impacting on the dust and gas radii. In this study we neglected these effects, that can be important for timescales longer than 3 Myr. Although unlikely to affect our results in Lupus, they are probably playing a role in Upper Sco. Dedicated studies are necessary to quantify the impact of these effects on discs sizes.

\subsubsection{Unresolved substructures}\label{sec:substr}

Without changing any assumption on the efficiency of the grain growth, on the initial conditions and on the viscous evolution of the gas (thus fixing the $R_{\rm CO}$ sizes), the only way to match the $R_{\rm CO}/R_{\rm dust}$ vs $R_{\rm dust}$ values is to find a mechanism able to slow down or stop radial drift, in order to enlarge (by at least a factor 2) the dust radii. Doubling the dust radius can also increase the match between the observed values of $R_{\rm CO}$ vs $R_{\rm dust}$ and our models, moving almost all the points from the $R_{\rm CO}=20 R_{\rm dust}$ to the $R_{\rm CO}\lesssim10 R_{\rm dust}$ without any change on the gas evolution. Substructures, already invoked to explain the large dust radii observed in bright discs, would in this case be at play in almost \textit{all} the discs, (e.g., also in the compact and faint ones), though in this case mostly undetected. We would thus predict that, if observed at high enough spatial resolution, even small and faint discs should show substructure in a similar way to large and bright discs. The ratio $R_{\rm CO}/R_{\rm dust}$ thus paints a similar picture as the measurements of spectral indices \citep{Testi2014}, which also require the presence of substructure to be explained \citep{Pinilla2012}.

Many theories can explain the formation of substructures. Hydro and magneto-hydro instabilities and interactions are capable to perturb the gas distribution, creating lcal pressure maxima. Examples are magnetic induced traps \citep{Suriano19}, photoevaporative flows \citep{Ercolano2017}, vortices, self-gravity \citep{Kratter2016}, dynamical interaction with detected or undetected companions \citep{Lin1979,Goldreich1980}; for a complete review see e.g., \citet{Andrews2020}. Other mechanisms are able to induce substructures in the gas distribution without local pressure maxima, e.g due to the sublimation of icy particles during migrations (snowlines, see e.g.,  \citealt{Stammler2017}). However, without pressure maxima, these mechanisms do not stop radial drift and would thus not solve the problem.

Any mechanism able to form a pressure maximum can stop or at least slow down radial drift, resulting in larger dust radii: at the location of the gas bumps, dust piles up, creating substructures such as rings \citep{Pinilla2012}. Dust grains drifting from the outer parts of the discs would stop drifting at the outermost bump location, resulting in a larger dust radius for the same gas radius, leading to smaller $R_{\rm CO}/R_{\rm dust}$ values \citep{Long2018}.  
In order to fix the dust radii on timescales of $\sim$ few Myr, dust traps have to be long lived: a natural mechanism fulfilling this request is the presence of planets interacting with the disc \citep{Goldreich1980}, while the question remains open on how long lived are structures created by the interaction with the magnetic field. On top of that, secondary production of dust can be at play, collisionally producing second-generation of dust \citep{Turrini2019}.

\subsubsection{Different chemical evolution}\label{sec:chemevo}
\begin{figure}
	\centering
	\includegraphics[width=0.75\linewidth]{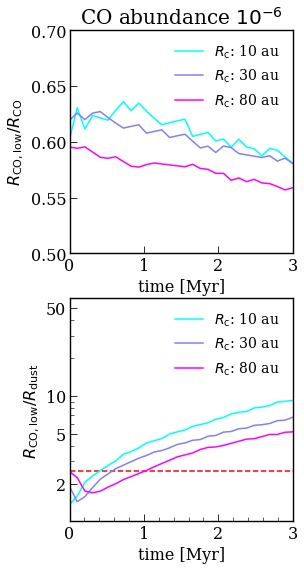}
    \caption{Top: Ratio between $R_{\rm CO, low}$ evaluated supposing a lower CO abundance ($10^{-6}$), and $R_{\rm CO}$. In this case, we show the reference case ($M_0=0.1$ M$_\odot$, $\alpha=10^{-3}$) with different values of $R_{\rm c}$ (cyan: $R_{\rm c}$=10~au, violet:  $R_{\rm c}$=30~au, magenta:  $R_{\rm c}$=80~au). Bottom: $R_{\rm CO, low}/R_{\rm dust}$ for the same models ($M_0=0.1$ M$_\odot$, $\alpha=10^{-3}$) with different values of $R_{\rm c}$ (cyan: $R_{\rm c}$=10~au, violet:  $R_{\rm c}$=30~au, magenta:  $R_{\rm c}$=80~au). The average value for the ratio $R_{\rm CO}/R_{\rm dust}$ in the Lupus population is also shown as a dashed red line.}
    \label{fig:mass_lowco}
\end{figure}
In this work we prescribed the standard abundances for the CO molecules, $10^{-4}$. However, several observations found lower abundances for CO (a factor 10-100 times smaller, see e.g., \citealt{Favre2013} for TW Hydra, \citealt{McClure2016} or \citealt{BerginWilliams2017} for an overview). Moreover, turbulence is expected to mix vertically and radially dust and gas in the disc. As a consequence, the upper layers of the disc, rich in CO, moves down towards the midplane, where the temperature is sufficiently low to allow the freeze-out of CO molecules onto the dust grains, where they can be converted in other species. In this work, we included a simple prescription for CO freeze out, but the presence of vertical mixing can increase the depletion of CO, decreasing the abundance of CO in the emitting layers, affecting the disc size \citep{Krjit2020}.
The gas radius depends on the CO content in the disc, while the dust radius, to a first approximation, does not (as the dynamics of the dust is dominated by the total gas mass); in particular, a smaller CO abundance leads to smaller radii \citepalias{Trapman2019} and, consequently, smaller $R_{\rm CO}/R_{\rm dust}$. To quantify how robust are our finding with respect to this parameter, we tested our reference models ($M_{0}=0.1$ M$_\odot, \alpha=10^{-3}$) for different values of the scale radius $R_{\rm c}$ (10, 30, 80 au), fixing the CO abundance to $10^{-6}$, a value among the lowest CO abundance inferred for discs in the Lupus star forming region.  Figure~\ref{fig:mass_lowco} shows the ratio between the gas radius evaluated in cubes with low CO abundance, $R_{\rm CO, low}$ and the standard case. Once again, our findings are in agreement with \citetalias{Trapman2019}, confirming that the chemistry of the $^{12}$CO is quite "simple": the lowest CO abundance leads to a smaller gas radius, about $\sim 60\%$ of the standard case. Thus,  prescribing a very low abundance for the CO can partially solve the problem of having large gas radii (assuming that all the radii are reduced to 60$\%$ of their size), but in order to explain the observed $R_{\rm CO}$/$R_{\rm dust}$ values (see also Figure~\ref{fig:mass_lowco}, bottom figure), a high depletion of CO should be considered (for a complete discussion see \citealt{Trapman2021}). To match the observed $R_{\rm CO}/R_{\rm dust}$, we can allow the 
the possible initial disc size to be twice the size of the standard case, $R_{\rm c} \sim 30-50$ au. However, the resulting dust radius is still small, implying also in this case the need of substructures in order to match the observed dust radii of the Lupus population (though the request could be less stringent than the standard case).  

\subsubsection{Different viscosity law}\label{sec:testgamma}
\begin{figure*}
	\centering
	\includegraphics[width=0.85\textwidth]{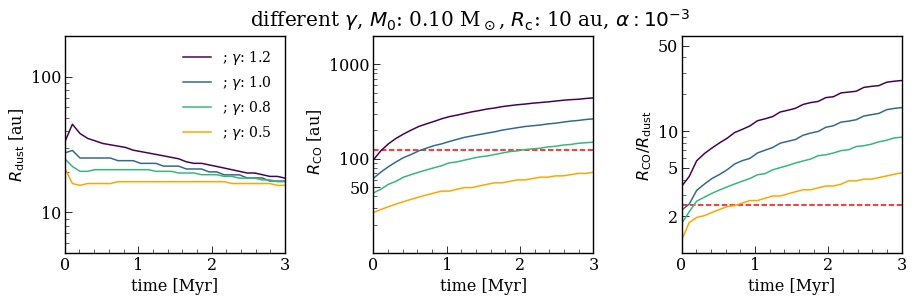}
    \caption{Dependence of the reference case on the value of  $\gamma$: left, middle and right panels shows the values of $R_{\rm dust}$, $R_{\rm CO}$ and the ratio $R_{\rm CO}/R_{\rm dust}$ as a function of disc age (in Myr) respectively, for different values of gamma (violet: $\gamma=1.2$, dark green:  $\gamma=1.0$, light green:  $\gamma=0.8$, yellow: $\gamma=0.5$) for discs with $\alpha = 10^{-3}$, $R_{\rm c}$=10~au and $M_0 = 0.1$~M$_\odot$. The average values for $R_{\rm CO}$ and the ratio $R_{\rm CO}/R_{\rm dust}$ in the Lupus population are also shown in  middle and right panels as dashed red lines.}
    \label{fig:fig_gamma}
\end{figure*}

Because the gas radii (both 68$\%$ and 90$\%$) are 5-10 times larger than the scale radius of the disc ($R_{\rm c}$), this implies that most of the mass (> 97$\%$ at 1 Myr) is enclosed in the gas radius \citepalias{Trapman2020}. 
As a result, the exact shape of outer part of the disc plays a major role in determining the gas radius. In this context, the choice of the viscosity law index, $\gamma$ is thus very important, because in the self-similar solution it also affects the sharpness of the outer exponential cut-off in the density, where lower values of $\gamma$ imply a sharper outer edge.
Thus, to take into account different truncation shapes of the outer part of the disc, we tested the impact of different values of $\gamma$ on the dust and gas radii. The results of the test are shown in Figure \ref{fig:fig_gamma}. Clearly, the value of the slope $\gamma$ has a strong impact in the dust and gas radii, that decrease for lower values of $\gamma$. However, after 2-3 Myr, due to the effect of radial drift, the value of $R_{\rm dust}$ is almost the same for all  values of $\gamma$, while the value of $R_{\rm CO}$ is affected similarly at all times. As a consequence, the value of $R_{\rm CO}/R_{\rm dust}$ is smaller for smaller values of $\gamma$. To help the reader, we reported in Fig. \ref{fig:fig_gamma} the mean values of $R_{\rm CO}$ and $R_{\rm CO}/R_{\rm dust}$ for the Lupus population, as red dashed lines. A value of $\gamma=0.5$ could significantly reduce the $R_{\rm CO}/R_{\rm dust}$ values,  allowing the initial disc size to be a factor 2 larger than in the standard case, $R_{\rm c} \simeq 30-50$ au. In a recent work, \citet{Dullemond2020} found that, in the case of the source HD163296, the surface density in the outer part may be steeper than the standard case, supporting this scenario (at least for this single source).
However, also in this case, the resulting dust radius is still small, requiring substructures to explain the observed values (again, the request could be less stringent than the standard case). Moreover, \citet{Lodato2017} found that values of $\gamma < 1$ fail to reproduce the observed accretion rates in Lupus region, and they require $\gamma\simeq 1.2$, which, in our model would lead to an even larger gas radius.

\section{Caveats}

\textit{Photoevaporation:} In this work we considered only pure viscous evolution of discs, that predicts a lifetime of $\sim$ few Myr (consistent with observational results, see e.g., \citealt{Fedele2010}). We did not consider any of the processes related to mass loss in the outer part of the disc and disc dispersal, such as external \citep{Facchini2016} and internal \citep{Clarke2001} photo-evaporation, as they are out of the scope of this paper. While internal photo-evaporation mostly affects the inner part of the disc, external photo-evaporation preferentially removes mass from the outer part of discs, potentially impacting disc radii. Indeed, \citealt{Sellek2020} found that external photo-evaporation can reduce the amount of dust in the outer part of the disc, increasing the efficiency of radial drift, thus reducing $R_{\rm dust}$. At the same time, also $R_{\rm CO}$ is expected to decrease if the accretion rate is smaller than the photo-evaporative mass loss rate. However, the importance of this effects depends on the star forming region: while in regions close to OB stars such as Upper Scorpius the effect is expected to be severe, in younger regions such as Lupus, far from high mass stars and exposed to low irradiation, the impact should be negligible  (at least for the bulk of the population, as large discs can be affected, for a broader discussion see e.g., \citetalias{Trapman2020}). To give an estimate of the different external radiation fields strength in the regions, according to \citetalias{Trapman2020} the external radiation field in Upper Scorpius should be $G_0 \sim 10 -300$ while \citet{Cleeves2016} find in IM Lup a value of $G_0 \lesssim 4$\footnote{A complete analysis of the radiation field in Lupus is still missing.}. However, we plan to include disc dispersal processes in our models in future works.

\subsection{Reliability of CO as a tracer}

\begin{figure}
	\centering
	\includegraphics[width=0.63\linewidth]{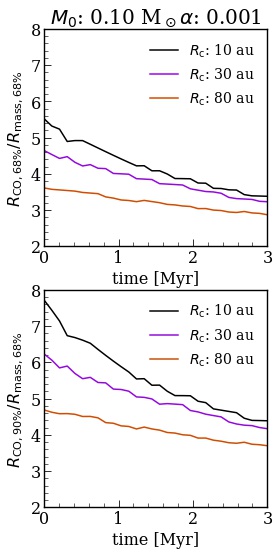}
    \caption{Ratio between $R_{\rm CO,68\%}$ and the radius that enclose the 68$\%$ of the disc mass, $R_{\rm mass, 68\%}$ (top) and between $R_{\rm CO,90\%}$ and $R_{\rm mass, 68\%}$ (bottom) as a function of the time for the fiducial model ($M_0=0.1$~M$_\odot$, and $\alpha=0.001$), considering different values of the initial disc size $R_{\rm c}$ (black: $R_{\rm c}=10$~au, purple:  $R_{\rm c}=30$~au, orange:  $R_{\rm c}=80$~au). As the figure points out, the gas radius is not a good mass and viscous spreading tracer.}
    \label{fig:mass_enclose}
\end{figure}

This work, \citetalias{Trapman2019} and \citetalias{Trapman2020} clearly show that $R_{\rm CO,68\%}$ and $R_{\rm CO,90\%}$ enclose most of the mass in the disc, as they are larger than several $R_{\rm c}$.  As we already showed in Sec. \ref{sec:testgamma}, this implies that the gas radius depends on the mass distribution in the outer part of the disc, described by the parameter $\gamma$ in the self-similar scenario. This hints that the measure of $R_{\rm CO,68\%}$ and $R_{\rm CO, 90\%}$ itself is not a good mass or viscous spreading tracer. To confirm this fact, we tested the time evolution of the ratio between the CO radius and the radius that encloses  68$\%$ of the disc mass, $R_{\rm mass,68\%}$. 
Figure \ref{fig:mass_enclose} shows the ratio between $R_{\rm CO,68\%}$ (top), $R_{\rm CO, 90\%}$ (bottom) and $R_{\rm mass, 68\%}$ as a function of time. Clearly, the ratio is not constant with time and has different evolution for different discs, implying that the $R_{\rm CO}$ is not a reliable measurement of the disc mass (or the scale radius $R_{\rm c}$) or the viscous spreading of the disc. Note that this plot presents the same information as \citetalias{Trapman2020}: again, our results are in agreement. We included the figure to foster the discussion.

As the gas radii probed with $^{12}$CO appear to be so dependent on the models, one may wonder if radii measured with different, optically thin molecules could be less sensitive to the caveats listed above. Indeed, in principle $^{13}$CO is expected to be a better tracer of the mass distribution as it probes the inner part of the disc and may not be affected by the uncertainties on $\gamma$ or by external photo-evaporation. This is due to the fact that its emission is more concentrated towards the inner disk, while the outer disk density is too low to significantly emit (all the $^{13}$CO is expected to be photodissociated there), making it less affected by the shape of the outer disc. However, it is largely dependent on the thermochemistry of the discs (see e.g., \citealt{Miotello2014,Trapman2021}) and the chemistry of the upper layers is better known, thus $^{12}$CO is our molecule of choice. However, as new data will be collected, the accuracy of thermochemical models will increase, and a detailed study of $^{13}$CO will be appropriate.

\section{Conclusions}\label{concl}

In this paper we modelled the time evolution of protoplanetary discs assuming viscous evolution, grain growth and radial drift. Pressure bumps and substructures are not included. After evaluating the 68$\%$ dust radius of the models as in \citetalias{Rosotti2019} and the 68$\%$ gas radius using the cumulative flux method on synthetic J=2-1 CO flux obtained with the radiative transfer code RADMC \citep{Dullemond2012}, we investigated how the ratio between the observed gas and dust radii of discs evolves as discs age, to test the role of radial drift in discs, focusing on its efficiency. 
Our main results are:
\begin{itemize}
    \item Radial drift has a strong impact on the evolution of dust radii, making them shrink quickly with time and resulting in high values of $R_{\rm CO}/R_{\rm dust}$.
    \item The values of dust radii $R_{\rm dust}$ in our models at 2~Myr are too small (< 40 au) and, consequently, the ratio $R_{\rm CO}/R_{\rm dust}$ is too large (>5) with respect to the observed values in Lupus \citep{Ansdell2018,Tazzari2020,Hendler2020,Sanchis2021}. The values of $R_{\rm CO}$ are compatible with observed values in Lupus (50-300 au).  
    \item Initially larger discs ($R_{\rm c}$ > 30 au) are unlikely: the resulting gas radii are too large with respect to the observed values and in contrast with observational results from class 0-I surveys VANDAM and CALYPSO \citep{Tobin2020,Maury2019}.
    \item We tested a reduced efficiency of radial drift, 
    finding that a strongly less effective grain growth, resulting in a less efficient radial drift and blocked disc evolution, could explain the observed larger values of $R_{\rm dust}$, lowering the $R_{\rm CO}/R_{\rm dust}$ values. However, this scenario would lead to extremely large disc masses according to \citealt{Sellek2020}.
    \item Consider disc wind driven accretion as the main driver of gas evolution allows to start with initially larger discs. However, in order to match the observed data, disc winds should efficiently remove mass from the disc, reducing the gas radii by a factor 2-4. This poses a very strong constraint on the efficiency of models, with  high accretion rates and winds to be observed. Moreover, this is in contrast with class 0-I surveys, thus this solution seems unlikely. 
    \item  The most likely hypothesis is the presence of unresolved or undetected substructures in \textit{most (or all)} discs. Dust grains drifting from the outer disk stop or slow down at the outer bump location \citep{Whipple1972,Pinilla2012}. This results in fixing the observed dust size $R_{\rm dust}$ to the outer edge of the outer dust trap, decreasing the $R_{\rm CO}/R_{\rm dust}$ size ratio regardless of the the mechanisms of angular momentum transport. In order for this explanation to work, these substructures have to be long lived rather than transient, such as pressure bumps created by planets.
    \item We tested the effect of reducing the  CO abundance in all the disc population to $10^{-6}$, finding that small values of disc viscosity ($\lesssim 10^{-3}-10^{-4}$), and initial disc size $R_{\rm c} \simeq 30$ au could match the observed gas radii. However, substructures are still needed in order to match the observed dust radii of the Lupus population (though the request could be less stringent than the standard case).
    \item As the distribution of gas in the outer part of the disc is crucial for the observed gas radius, we considered the impact of different viscosity laws, finding that $\gamma \lesssim 0.5$ could reduce $R_{\rm CO}$ without reducing $R_{\rm dust}$. In this case, larger initial disc sizes are possible, $R_{\rm c} \simeq 30-50$ au. However, $\gamma \lesssim 0.5$ fails to reproduce the observed mass accretion rates in the Lupus region, that indeed requires  $\gamma \gtrsim 1.2$ \citep{Lodato2017}. 
\end{itemize}

In this work we focus on discs where we can detect $R_{\rm CO}$ and $R_{\rm dust}$, but an important open question is to understand what happens in faint and compact sources ($R<1$ au). It is possible that in these discs all the large grain did indeed drift and they became sterile discs that will never form planets.
For these sources also gas radii measurements lack, but if this is the case, our models predict $R_{\rm CO}/R_{\rm dust}$ values larger than 4. 

We plan to perform future studies using a population synthesis approach, to improve our results. Different values of initial stellar mass and an appropriate set of initial conditions will allow us to tailor our model to the different star forming region, improving the quality of our comparisons with observations and extending our findings also to other star forming regions.

\section{Data availability}
All the datasets generated and analysed during this study are available from the corresponding author on reasonable request.

\section*{Acknowledgements}
The authors thank the reviewer Richard Booth for his thoughtful comments towards improving our manuscript. This project, CT and GL have received funding from the European Union's Horizon 2020 research and innovation programme under the Marie Skłodowska-Curie grant agreement No 823823 (DUSTBUSTERS RISE project). GR acknowledges support from the Netherlands Organisation for Scientific Research (NWO, program number 016.Veni.192.233) and from an STFC Ernest Rutherford Fellowship (grant number ST/T003855/1). LTr acknowledges the support of the Office of the Vice Chancellor for Research and Graduate Education at the University of Wisconsin – Madison with funding from the Wisconsin Alumni Research Foundation and NWO grant 614.001.352.
LTe acknowledges support from Italian Ministry of Education, Universities and Research through the grant Progetti Premiali 2012 iALMA(CUP C52I13000140001), the Deutsche Forschungsgemeinschaft (DFG, German Research Foundation) - Ref no. FOR 2634/1ER685/11-
1 and the DFG cluster of excellence ORIGINS (www.originscluster.de), and the European Research Council (ERC)
via the ERC Synergy Grant ECOGAL (grant 855130). 
This work made use of the Python packages Numpy, matplotlib, Pandas and Seaborn.




\bibliographystyle{mnras}
\bibliography{mnras_template} 



\appendix

\section{Comparison with previous models}\label{App_1}

To test if the synthetic CO flux we obtained using the RADMC code, with our parametric description of CO freeze-out and photodissociation, assuming only the presence of CO molecules, is a good estimate of the emitted flux of discs with more complex chemistry, we compare our results with the ones obtained by \citetalias{Trapman2020}. In their work, the authors used the thermochemical code DALI, that includes a more complex chemistry compared with our hypothesis (see Sec.~2.3 of their paper), to estimate the thermal and chemical structure of the disc, creating synthetic emission maps of CO molecules, evaluating with the cumulative flux method the $R_{\rm CO,90}$ as the radius that encloses 90$\%$ of the CO J=2-1 flux. They fixed the accretion rates of the star and the initial disc size $R_{\rm c}=10$ au, varying the disc masses for different values of $\alpha$.

\begin{figure}
	\centering
	\includegraphics[width=0.7\linewidth]{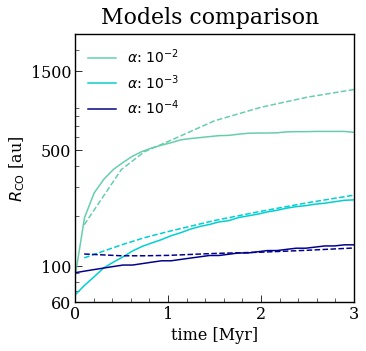}
    \caption{$R_{\rm CO}$ as a function of disc age (in Myr) for models with different viscosity ($\alpha=10^{-4}$,$10^{-3}$,$10^{-2}$), initial disc masses (0.069 M$_\odot$ for $\alpha=10^{-2}$, 0.059 M$_\odot$ for $\alpha=10^{-3}$ and 0.26 M$_\odot$ for $\alpha=10^{-4}$) and initial radius $R_{\rm c}=10$ au (solid line), to compare with the same models from \citetalias{Trapman2020} (dashed lines).}
    \label{fig:A_1}
\end{figure}

We show in Fig.~\ref{fig:A_1} a comparison between the values of the gas radii as a function of disc age (in Myr) for models that have the same initial conditions of \citetalias{Trapman2020} (initial disc masses of 0.069 M$_\odot$ for $\alpha=10^{-2}$, 0.059 M$_\odot$ for $\alpha=10^{-3}$ and 0.26 M$_\odot$ for $\alpha=10^{-4}$) and the results from \citetalias{Trapman2020}, represented with solid and dashed lines respectively. 
To properly compare the two results, we computed $R_{\rm CO,90}$ of the CO J=2-1 flux for our models. The values of $R_{\rm CO,90}$ shown in Fig.~\ref{App_1} are the ones shown in Fig. 3 (bottom panel) of \citetalias{Trapman2020}, and are obtained considering a star of 1.0 M$_\odot$. Despite the numerical differences given by the different codes, we can conclude that also our estimates are reliable. 
Note that current observations suggest that the CO abundance decreases over Myr timescales (see, e.g., \citep{Miotello2017,Zhang2020}. It has been suggested that CO is either chemically converted on the surface of dust grains into CO$_2$ and more complex molecules \citep[e.g.,]{Aikawa1997,Bosman2018}, or it is being locked up in larger dust bodies \citep[e.g.,]{Kama2016, Krijt2018} 
However, it is currently unclear how these processes affect the observed gas disk size (see \citetalias{Trapman2020}, sect 4.3). Moreover, internal and external photoevaporation are expected to play a role for longer timescales, and may have an influence on the final gas radii extension.

\section{Detailed scatter plot of the models}\label{App_2}
\begin{figure*}
	\centering
	\includegraphics[width=0.95\textwidth]{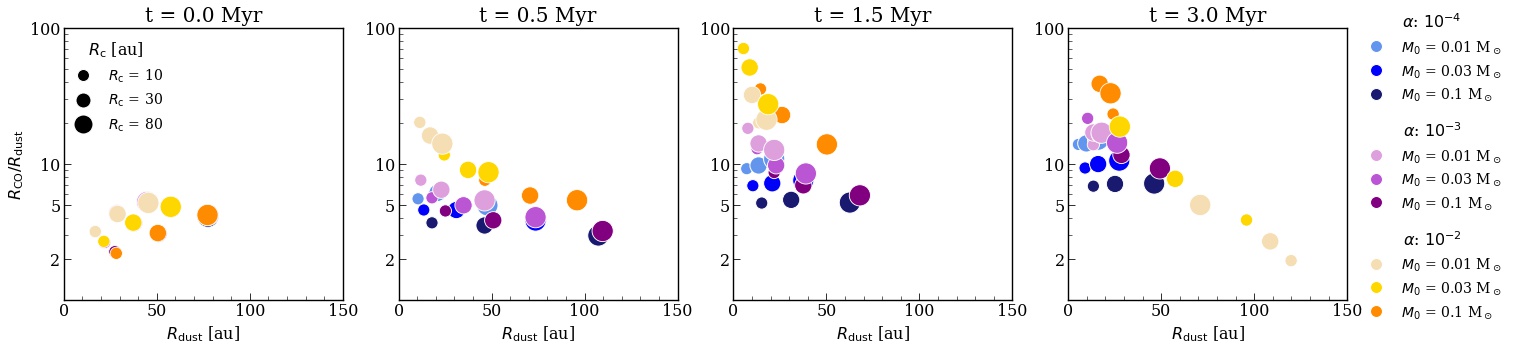}
    \caption{Four different snapshots of $R_{\rm CO}/R_{\rm dust}$ vs $R_{\rm dust}$ (in au) at 0 (left), 0.5 and 1.5 (middle) and 3 (right) Myr respectively. Different values of viscosity are shown with different palettes of colours ($\alpha=10^{-4}$ blue, $\alpha=10^{-3}$ purple, $\alpha=10^{-2}$ orange), different initial sizes are represented with different circle sizes (larger for initially wider disc sizes) and different masses are displayed with different tones of the colours (darker for initially more massive discs). }
    \label{fig:A_3}
\end{figure*}
\begin{figure*}
	\centering
	\includegraphics[width=0.95\textwidth]{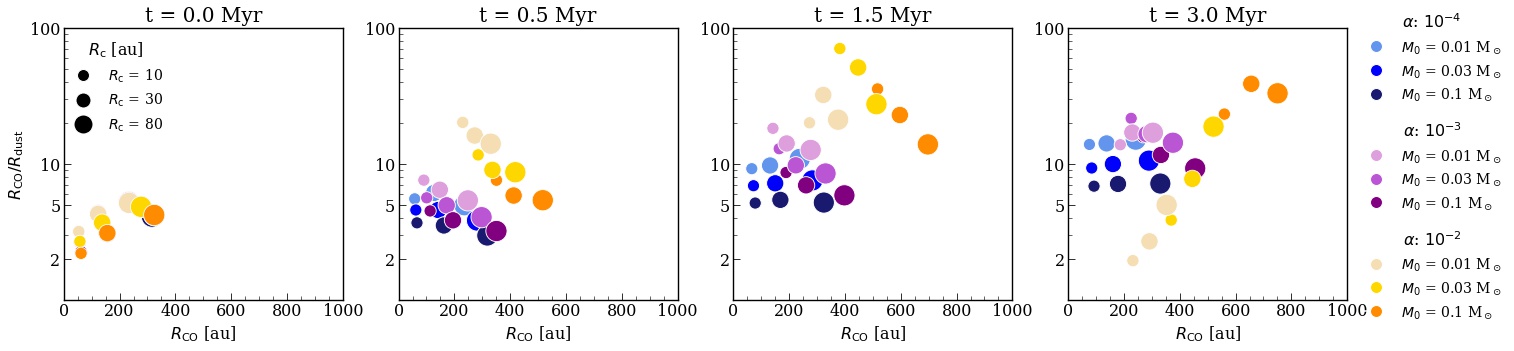}
    \caption{Three different snapshots of $R_{\rm CO}/R_{\rm dust}$ vs $R_{\rm CO}$ (in au) at 0 (left), 0.5 and 1.5 (middle) and 3 (right) Myr respectively. Different values of viscosity are shown with different palettes of colours ($\alpha=10^{-4}$ blue, $\alpha=10^{-3}$ purple, $\alpha=10^{-2}$ orange), different initial sizes are represented with different circle sizes (larger for initially wider disc sizes) and different masses are displayed with different tones of the colours (darker for initially more massive discs). }
    \label{fig:A_4}
\end{figure*}
Fig.~\ref{fig:A_3} show the four scatter plots representing the snapshots of $R_{\rm CO}/R_{\rm dust}$ vs $R_{\rm dust}$ (in au) at 0 (left), 0.5 and 1.5 (middle) and 3 (right) Myr respectively. All the models are marked to understand their initial conditions. After 0.5~Myr (middle panel) all the models have $R_{\rm CO}/R_{\rm dust} \gtrsim 3$, larger for higher values of the viscosity. The values of gas and dust radii of these discs is still very dependent on the initial conditions: for very young population of discs (i.e. class 0 and early class I sources), dust is still growing, thus radial drift is expected to play a minor role in the observed dust radius: $R_{\rm dust}$ can be very large, > 50 au, for initially larger discs. This is in agreement with theoretical models \citep{Birnstiel2012}, where radial drift is expected to play a key role after 1 Myr, as well as with observations of early sources \citep{Maury2019,Tobin2020}
After 1.5~Myr (middle panel) all the models have $R_{\rm CO}/R_{\rm dust}$ > than 4, larger for higher values of the viscosity. Larger and heavier discs have smaller values of $R_{\rm CO}/R_{\rm dust}$ and larger values of $R_{\rm dust}$, $\sim$ 10 au. Initially smaller and lighter discs have very small values of  $R_{\rm dust}$, < 30 au. After 3~Myr (right panel), discs with high viscosity, compact and with smaller masses have $R_{\rm CO}/R_{\rm dust}$ $\lesssim$ 5, with very large values of $R_{\rm dust}$ (> 60 au) while all the other discs have $R_{\rm dust}$ < 50~au  and $R_{\rm CO}/R_{\rm dust}$ $\gtrsim$ 5. In particular, smaller and lighter discs have larger $R_{\rm CO}/R_{\rm dust}$.

Fig.~\ref{fig:A_4} displays $R_{\rm CO}/R_{\rm dust}$ vs $R_{\rm CO}$ for discs with the same age of Fig.~\ref{fig:A_3}, with the same legend as before. After 1.5~Myr (middle panel) smaller and lighter discs have smaller values of $R_{\rm CO}$; discs with $\alpha=10^{-2}$ have the wider values of $R_{\rm CO}$, $\gtrsim$ 400, larger for initially more massive discs, while discs with $\alpha=10^{-4}$ have more compact gas radii (< 400 au). After 3~Myr (right panel), all the discs with small and moderate values of viscosity still have  $R_{\rm CO}$ < 600~au and $R_{\rm CO}/R_{\rm dust}$ $\sim 5-20$, while discs with high viscosity are split in two different parts of the plot: initially compact and lighter discs have $R_{\rm CO}/R_{\rm dust}$ $\lesssim$ 5 and values of $R_{\rm CO}$ smaller than 500 au, meaning that they are reducing their observed gas radius, while more extended and heavier discs have  values of $R_{\rm CO}$ larger than at 1.5 Myr, $\gtrsim$ 700 au. 

\section{Effect of different values of the flux threshold}\label{App_3}
\begin{figure*}
	\centering
	\includegraphics[width=0.7\textwidth]{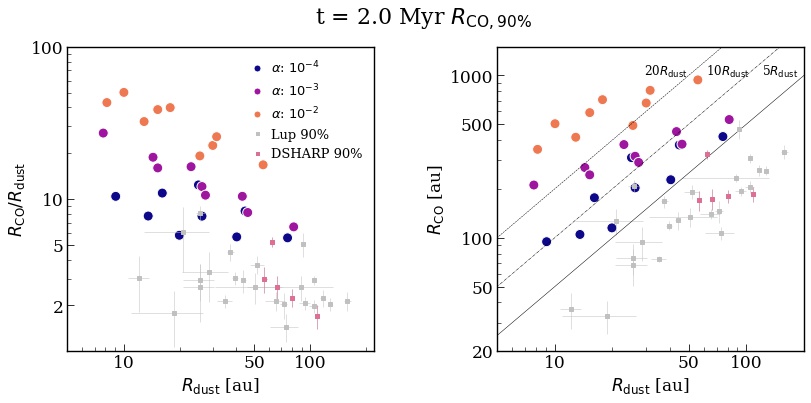}
    \caption{Comparison between the models at 2~Myr (dots) and the observed values in Lupus of $R_{\rm CO,90\%}/R_{\rm dust,90\%}$ (top) and $R_{\rm CO,90\%}$ (in au, bottom) vs $R_{\rm dust,90\%}$ (in au). Different values of viscosity are shown with different colours ($\alpha=10^{-4}$ blue, $\alpha=10^{-3}$ purple, $\alpha=10^{-2}$ orange). $R_{\rm CO,90\%}$ and $R_{\rm dust,90\%}$ values in Lupus shown in light grey are from \citealt{Sanchis2021}, the pink squares are the targets observed also in the DSHARP survey \citep{Andrews2018}. In the bottom figure are also shown the lines $R_{\rm CO,90\%}= 5,10,15 R_{\rm dust,90\%}$ with solid, dashed and dotted lines respectively. }
    \label{fig:A_5}
\end{figure*}
The flux cutoffs used in the cumulative flux analysis can significantly affect the estimate of the radius, especially in the outer part of the disc, where emission can be very faint. Moreover, for gas observations signal to noise can be very low \citep{Ansdell2018}.
To test if different values of the flux threshold leads to different conclusions in our analysis, we show in Fig. \ref{fig:A_5} the comparison between our models and Lupus data from \citet{Sanchis2021}, \citet{Hendler2020} and \citet{Ansdell2018} for dust and gas radii defined as the radius that contains 90$\%$ of the flux of the given tracer, namely $R_{\rm CO,90\%}$ and $R_{\rm dust,90\%}$. As in \citetalias{Rosotti2019}, we considered in our $R_{\rm dust,90\%}$ estimate the impact of the finite telescope sensitivity of real observations (sensitivity of 6 $\times 10^7$ Jy/sr as in standard ALMA surveys).
While the dust and gas radii of the single source increases, the behaviour of the disc population in Lupus shows no significant difference with respect to considering $R_{\rm CO,68\%}$ and $R_{\rm dust,68\%}$.


\bsp	
\label{lastpage}
\end{document}